\documentclass[10pt,letterpaper]{article}
\usepackage{preamble}

\usepackage[top=0.85in,left=2.75in,footskip=0.75in]{geometry}

\usepackage{amsmath,amssymb}

\usepackage{changepage}

\usepackage[utf8x]{inputenc}

\usepackage{textcomp,marvosym}

\usepackage{cite}

\usepackage{nameref,hyperref}

\usepackage[right]{lineno}

\usepackage{microtype}
\DisableLigatures[f]{encoding = *, family = * }

\usepackage[table]{xcolor}

\usepackage{array}

\newcolumntype{+}{!{\vrule width 2pt}}

\newlength\savedwidth

\newcommand\thickhline{\noalign{\global\savedwidth\arrayrulewidth\global\arrayrulewidth 2pt}%
\hline
\noalign{\global\arrayrulewidth\savedwidth}}

\raggedright
\usepackage[aboveskip=1pt,labelfont=bf,labelsep=period,justification=raggedright,singlelinecheck=off]{caption}

\makeatletter
\renewcommand{\@biblabel}[1]{\quad#1.}
\makeatother

\usepackage{lastpage,fancyhdr,graphicx}
\usepackage{epstopdf}
\fancyheadoffset[L]{2.25in}
\fancyfootoffset[L]{2.25in}
\begin{document}
\vspace*{0.2in}
\graphicspath{ {figures/} }

\begin{flushleft}
{\Large
\textbf\newline{Estimating fine age structure and time trends in human contact patterns from coarse contact data: the Bayesian rate consistency model}
}
\newline
\\
Shozen Dan\textsuperscript{1$\dagger$\Yinyang*},
Yu Chen\textsuperscript{1$\dagger$\Yinyang},
Yining Chen\textsuperscript{1},
Melodie Monod\textsuperscript{1}, 
Veronika K Jaeger\textsuperscript{2}, 
Samir Bhatt\textsuperscript{3,4},
André Karch\textsuperscript{2},
Oliver Ratmann\textsuperscript{1\Yinyang*}
on behalf of the Machine Learning \& Global Health network
\\

\bigskip
\textbf{1} Department of Mathematics, Imperial College London, England, United Kingdom\\
\textbf{2} Institute of Epidemiology and Social Medicine, University of Münster, Germany \\
\textbf{3} School of Public Health, Imperial College London, England, United Kingdom\\
\textbf{4} Department of Public Health, University of Copenhagen, Denmark
\bigskip

%
%
$\dagger$ Joint first authors.\\
\Yinyang\: These authors contributed equally to this work.\\


\textcurrency Current Address: Imperial College London, Exhibition Road, London SW7 2AZ, United Kingdom 

* Corresponding author: Shozen Dan, shozen.dan21@imperial.ac.uk, Oliver Ratmann, oliver.ratmann@imperial.ac.uk

\end{flushleft}
\section*{Abstract}
Since the emergence of severe acute respiratory syndrome coronavirus 2 (SARS-CoV-2), many contact surveys have been conducted to measure the fundamental changes in human interactions that occurred in the face of the pandemic and non-pharmaceutical interventions. These surveys were typically conducted longitudinally, using protocols that have important differences from those used in the pre-pandemic era. 
Here, we present a model-based statistical approach that can reconstruct contact patterns at 1-year resolution even when the age of the contacts is reported coarsely by 5 or 10-year age bands. This innovation is rooted in population-level consistency constraints in how contacts between groups must add up, which prompts us to call the approach presented here the Bayesian rate consistency model. The model also incorporates computationally efficient Hilbert Space Gaussian process priors to infer the dynamics in age- and gender-structured social contacts, and is designed to adjust for reporting fatigue emerging in longitudinal surveys.

On simulations, we show that social contact patterns by gender and 1-year age interval can indeed be reconstructed with adequate accuracy from coarsely reported data and within a fully Bayesian framework to quantify uncertainty. We then investigate the patterns and dynamics of social contact data collected in Germany from April to June 2020 across five longitudinal survey waves. We estimate the fine age structure in social contacts during the early stages of the pandemic and demonstrate that social contact intensities rebounded in a structured, non-homogeneous manner. We also show that by July 2020, social contact intensities remained well below pre-pandemic values despite a considerable easing of non-pharmaceutical interventions.

The Bayesian rate consistency model provides a modern, non-parametric, computationally tractable approach for estimating the fine structure and longitudinal trends in social contacts, and is readily applicable to contemporary survey data as long as the exact age of survey participants is reported.

\section*{Author summary}
The transmission of respiratory infectious diseases occurs during close social contacts. Hence, characterising social contact patterns within a population, encoded in contact matrices, leads to a better understanding of disease spread. Contact matrices also parameterise mathematical models, which played a pivotal role in informing health policy during the coronavirus disease 2019 (COVID-19) pandemic. Unlike pre-pandemic surveys, which recorded contacts' age in one-year age intervals, COVID-era studies recorded contacts' age in discrete large age categories to facilitate reporting. Some studies allowed participants to report an estimate for the total number of contacts for which they could not remember age and gender information. Many studies were partially longitudinal, which introduced the issue of reporting fatigue. Thus, directly applying existing statistical methods for estimating social contact matrices may result in a loss of age detail and confounded estimates. To this end, we develop a longitudinal model-based approach which estimates fine-age contact patterns from coarse-age data that also adjusts for the confounding effects of aggregate contact reporting and reporting fatigue in a unified manner. We apply our approach to the COVIMOD study to provide a detailed picture of how social contact dynamics evolved during the first wave of COVID-19 in Germany.

\nolinenumbers

\section*{Introduction}
The transmission of human respiratory diseases such as influenza, tuberculosis, and COVID-19 is directly driven by the rate of close social contact between individuals. Social contact studies such as the pivotal POLYMOD study~\cite{mossong_social_2008} have been widely acknowledged as an effective method of obtaining social contact estimates to assess infection risk and to parameterise mathematical infectious disease models~\cite{goeyvaerts_estimating_2010, eichner_4flu_2014, schmidt-ott_influence_2016}. Consequently, they provide critical epidemiological insights which inform the implementation and evaluation of non-pharmaceutical interventions~\cite{leung_transmissibility_2021} as well as public health policies such as vaccination schedules~\cite{wallinga_optimizing_2010}.

Since the outbreak of COVID-19, numerous contact studies have been conducted in Europe and around the world, providing indispensable information on the evolving patterns of human mixing behaviour during the pandemic \cite{verelst_socrates-comix_2021, feehan_quantifying_2021}. In Germany, the COVIMOD study collected social contact data for close to two years, and initial analyses~\cite{tomori_individual_2021} focused on data from April to June 2020 during the first partial lockdown in Germany to quantify the scale of social contact reductions relative to pre-pandemic contact patterns. High-resolution estimates of age- and gender-specific human contact patterns and their trends over time are substantially more difficult to obtain, due to limitations in available inference methods~\cite{goeyvaerts_estimating_2010, van_de_kassteele_efficient_2017, funk_socialmixr_2020}.

First, COVIMOD and other COVID-era social contact studies record the age of contacts by large age categories of 5 to 10 years, reflecting that often it is challenging for study participants to know the exact age of their contacts. In contrast, the pre-pandemic POLYMOD surveys collected data on the exact age of contacts and subsequently developed methods including thin-plate regressions~\cite{goeyvaerts_estimating_2010} and Gaussian Markov Random Field model approaches~\cite{van_de_kassteele_efficient_2017} relied on such data to estimate high-resolution contact patterns. The bootstrap approach implemented in the \texttt{socialmixr} library~\cite{funk_socialmixr_2020} provides a convenient and speedy way to estimate contact matrices. But, it only provides contact estimates in discrete large age categories, which is unsatisfactory as it may mask subtle but important age effects~\cite{farrington_contact_2005}.

Second, most COVID-era studies adopted retrospective web-based survey protocols and conducted longitudinal repeat surveys~\cite{gimma_changes_2022, coletti_comix_2020, backer_impact_2021, verelst_socrates-comix_2021}. Importantly, the survey waves are typically inter-dependent because a varying number of participants were surveyed in multiple waves, and additional participants were recruited to replenish the cohort size. While this approach provides valuable longitudinal data, it also introduces the issue of reporting fatigue, where participants tend to report fewer contacts in subsequent participation due to becoming tired of filling out the survey. It follows that directly applying existing methods, which do not incorporate adjustments to counter the confounding reporting fatigue effects, is bound to lead to incorrect estimates. Additionally, participants in the COVID-era surveys sometimes found it difficult to recall specific age and gender information for all of their contacts. Instead, they were allowed to report an estimate for the total number of contacts on that occasion~\cite{tomori_individual_2021}, which again may result in under-ascertainment of contact intensities if these data are not accounted for in inference approaches of contact patterns.

In this work, we present a temporal Bayesian model to infer age- and gender-specific contact patterns and trends at high 1-year resolution from longitudinal survey data. The primary innovation of the model is the ability to infer contact patterns by 1-year age bands even when the age contacts are reported in broad age categories. We call this model-based approach the Bayesian rate consistency model for reasons that will be clear soon. In addition, we use recently developed Hilbert Space Gaussian Process approximations~\cite{solin_hilbert_2020} to gain substantial advances in computational efficiency, which in turn enable us to make full Bayesian inferences over time and uncover the dynamics in social contact structure. We demonstrate that it is crucial to model contact patterns over time to account for reporting fatigue effects in inter-dependent longitudinal survey waves. The primary purpose behind developing the Bayesian rate consistency model is its application to contemporary COVID19-era survey data, which we present for data spanning the first five survey waves of the COVIMOD study in Germany. We present high-resolution estimates of age- and gender-specific social contacts for each survey wave and describe their time evolution. We also place the inferred contact dynamics into a pre-pandemic context and quantify the differences in contact intensity change by the age of contacts. 


\section*{Methods}
\subsection*{The COVIMOD study}
The COVIMOD study was launched in April 2020 and continued until December 2021, constituting 33 survey waves. Participants were recruited through email invitations to existing panel members of the online market research platform IPSOS i-say~\cite{ipsos_ipsos_2022}. To ensure the sample's broad representativeness of the German population, quota sampling was conducted based on age, gender, and region. Participants were invited to participate in multiple waves to track changes in social behaviour and attitudes toward COVID-19. When the participant size did not meet the sampling quota due to study withdrawals, new participants were recruited into the study. This approach enabled COVIMOD study to obtain longitudinal samples, but it also introduced the issue of response fatigue, where the number of detailed contacts reported decreased compared to previous participation, irrespective of the survey wave. To procure information on children, a subgroup of adult participants living with children under the age of 18 were selected to be proxies. This procedure meant that middle-aged adults were under-sampled as they completed the survey on behalf of their children.

The COVIMOD questionnaire was based on the CoMix study and includes questions on demographics, the presence of a household member belonging to a high-risk group, attitudes towards COVID-19 as well as related government measures, and current preventative behaviors~\cite{jarvis_quantifying_2020, tomori_individual_2021}. Participants were also asked to provide information about their social contacts between 5 a.m. the preceding day to 5 a.m. the day of answering the survey. Following the pre-pandemic POLYMOD study, a contact is defined as either a skin-to-skin contact such as a kiss or a handshake (physical contact) or an exchange of words in the presence of another person (non-physical contacts)~\cite{mossong_social_2008}. Participants were asked to report the age group, gender, relation, the contact setting (e.g. home, school, workplace, place of entertainment, etc.), and whether the contact was a household member. For survey waves 1 and 2, participants were asked to provide each contact's information separately. However, some participants reported contacts to groups of individuals (e.g., customers, clients) for which a specific number of contacts was assumed (Additional file 2 of~\cite{tomori_individual_2021}). From wave 3 onwards, participants were given the possibility to record group contacts in addition to the recording of individual contacts.
Additionally, some participants could not recall or preferred not to answer the age or gender information of some individual contacts. We treat these three types of entries with missing age or gender equally and refer to them as \emph{missing} \& \emph{aggregate contact reports}. A copy of the COVIMOD questionnaire may be found in Additional file 1 of~\cite{tomori_individual_2021}. COVIMOD was approved by the ethics committee of the Medical Board Westfalen-Lippe and the University of Münster, reference number 2020-473-f-s.

\begin{figure}[!t]
\begin{adjustwidth}{-2.25in}{0in}
    \includegraphics[width=\linewidth]{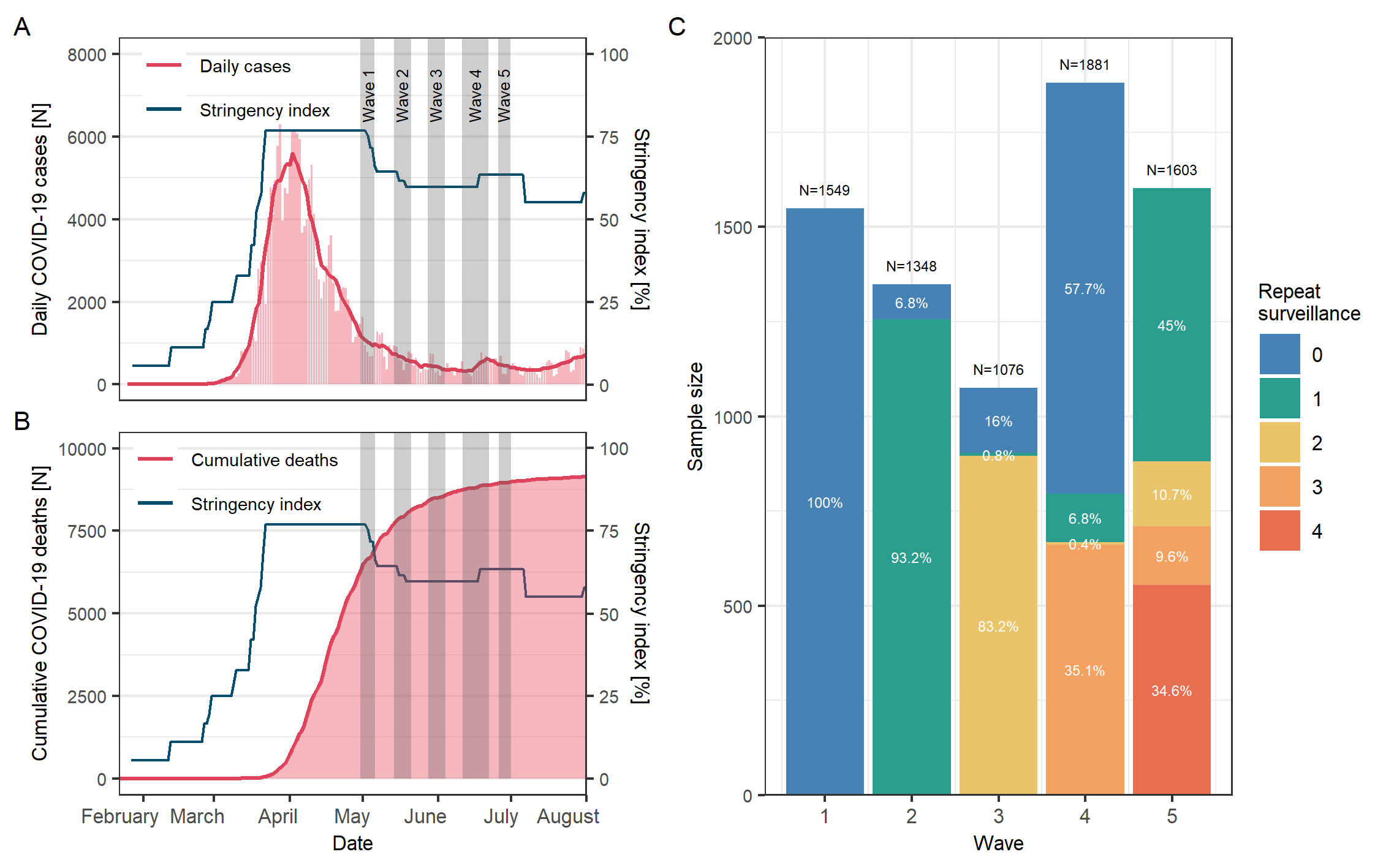}
    \begin{adjustwidth}{0.75in}{0in}
        \captionsetup{width=\linewidth}
        \caption{{\bf Timeline and participants of the longitudinal COVIMOD study} A. Daily COVID-19 case counts in Germany (red bars), the OxCGRT Stringency Index (blue line), and COVIMOD survey administration periods (grey ribbons). B. Cumulative COVID-19-related deaths in Germany (red line), the OxCGRT Stringency Index (blue line), and COVIMOD survey administration periods (grey ribbons). C. Sample sizes and the proportion of people repeatedly sampled in the COVIMOD survey for which zero repeats indicate first-time participants.}
        \label{fig:1}
    \end{adjustwidth}
\end{adjustwidth}
\end{figure}

This current work concerns the first five survey waves of the COVIMOD study. In Fig~\ref{fig:1}A and B, we show the sampling periods with the number of daily COVID-19 cases, cumulative COVID-19-related deaths, and the OxCGRT Stringency Index~\cite{hale_global_2021}. The following COVID-19 policy timeline is obtained from the ACAPS COVID-19 Government Measures dataset~\cite{acaps_covid}. The first COVIMOD survey was administered from April 30\textsuperscript{th} to May 6\textsuperscript{th} in year 2020, towards the end of the first partial lockdown and the first wave of cases. Before the beginning of the first survey (April 20\textsuperscript{th}), small stores, auto dealers, and bookstores were allowed to reopen under strict hygiene regulations. During the final few days of the survey period (May 4\textsuperscript{th} to 6\textsuperscript{th}), phase-out measures were announced by the government, including the step-wise uptake of schools, the reopening of hairdressers under strict hygiene regulations, lifting of the ban on public gatherings of 30 people indoors and 50 outdoors, resumption of religious services, and reopening of public services such as museums, botanical gardens, zoos, and playgrounds. The second wave of the COVIMOD survey was administered from May 14\textsuperscript{th} to May 21\textsuperscript{st}. During this period, additional phase-out measures were announced, including the resumption of all cross-country transport and the reopening of hotels and restaurants. International travel to neighbouring countries was also slightly relaxed during this period. The third, fourth, and fifth waves of COVIMOD surveys were taken from May 28\textsuperscript{th} to July 4\textsuperscript{th}, June 11\textsuperscript{th}-22\textsuperscript{nd}, and June 26\textsuperscript{th} to July 1\textsuperscript{st}, respectively. There was no notable introduction or reduction of social contact restriction measures during this time, but international travel restrictions were relaxed primarily for Schengen and EU countries. COVID-19 cases and deaths remained stable during this period (Fig~\ref{fig:1}).

After excluding participants who prefer not to provide age or gender information and 25 participants above the age of 84, there were 1549, 1345, 1076, 1881, and 1603 participants for waves 1 to 5. We observed 3244, 4852, 6344, 13471, and 8353 total contacts for each wave. In Fig~\ref{fig:1}C, we show the proportion of participants who consented to the survey multiple times. Most participants in waves 2 and 3 had participated in wave 1, with only 6.8\% and 16\% of participants being new to the survey. The proportion dropped sharply in wave 4, where only 35.1\% of initial participants remained. Hence the majority (57.7\%) of wave 4 participants were first-time participants. On the contrary, no new participants were enrolled for wave 5, and individuals who participated for the second, third, and fifth time took up approximately 45\%, 10.7\%, 9.6\%, and 34.6\% of the sample.   

%
%
%
%
%
%

\subsection*{Data processing}
Following ethical guidelines, the participant age information for children is reported in discrete categories, i.e., $\text{0−4, 5−9, 10−14, 15−18}$. To obtain fine-age information for participants
under 18, we imputed their age by drawing from a discrete uniform distribution with
bounds set as the minimum and maximum age of the participant’s age category. We
excluded 20 participants (0.3\% of the total) without age or gender information as we
could not estimate the information accurately. For total contact counts $Y^{gh}_{ab}$ between all participants aged $a \in \{0, 1, \ldots, 84\}$ of gender $g \in \{M, F\}$ and contacted individuals aged $b \in \{0, 1, \ldots, 84\}$ of gender $h \in \{M, F\}$, we filled in
missing entries with zeroes if a participant of age $a$ and gender $g$ is present. If not, we treated the entries as missing. We truncated group contacts at 60 (90\textsuperscript{th} percentile of group contacts) to remove the effects of extreme outliers. We show the distribution of observed contacts in \nameref{S1_Fig}.

%
%
%
%
%
%

\subsection*{Estimating contact patterns}
Throughout, we focus on estimating contact patterns and dynamics between men and women (superscripts $g, h \in \{M, F\}$) and high resolution one year age groups (subscripts $a, b \in \mathcal{B} = \{0, 1, 2, \cdots\, 84\}$). To introduce basic notations, let us momentarily consider male-to-female contacts at some fixed time $t$ and suppress the time index. We start by considering the total number of contacts $\ccntMF$ that are reported by men of age $a$ who participate in the survey to all women of age $b$ in the population. The number of male survey participants of age $a$ is $\PartM$, and the number of women of age $b$ in the population is $\PopF$. When $\PartM>0$ for age group $a$, the contact counts $\ccntMF$ are defined for all $b \in \mathcal{B}$, and are either zero or positive. When there are no participants for some age group $a$, there is no corresponding contact data, and we denote the participant age groups in the survey by $\AgesPartM = \{ a \in \mathcal{B} \colon \PartM>0\}$. Finally, we denote the number of age groups in $\AgesPartM$ by $A^M$ and the number of age groups in $\mathcal{B}$ by $B$. We also consider analogous notations for all other gender combinations using the superscripts $g$ and $h$. 

From the observed data, we seek to estimate the \emph{contact intensity} $\cintMF$, the average number of contacts from one male participant of age $a$ to all women aged $b$ in the population. We follow~\cite{mossong_social_2008, hens_mining_2009, goeyvaerts_estimating_2010, van_de_kassteele_efficient_2017, vandendijck_cohort-based_2022, wallinga_using_2006, feehan_quantifying_2021} and model the count data with an overdispersion adjusting Negative Binomial observation model, in shape-scale form,
\begin{subequations}\label{lkl-contact-patterns}
\begin{align}
\ccnt & \sim \text{NegBinomial}\left(\alpha_{ab}^{gh}, \frac{\odisp}{1+\odisp}\right)\\
\eccnt & = \alpha_{ab}^{gh}\odisp\\
\log \eccnt & = \log \cint + \log \Part \label{lkl-contact-patterns-b}
\end{align}
\end{subequations}
where $g,h\in\{M,F\}$, $a\in\AgesPartG$, $b\in\mathcal{B}$, and $\eccnt$ are the expected contacts that are expressed of the target quantity of interest, $\cint$, and the known $\Part$. The overdispersion parameter $\odisp>0$ such that $\text{Var}[Y_{ab}^{gh}] = \mu_{ab}^{gh}(1 + \nu)$ is allowed to be greater than the mean. 

The \emph{contact rate} is defined as the probability of contact between one male aged $a$ and one female aged $b$, i.e.
\begin{equation}\label{contact-rate}
    \crateMF = \cintMF / \PopF.
\end{equation}
Crucially, in a closed population, the (unknown) number of total contacts must be self-consistent,~i.~e.
\begin{equation}\label{contact-rate-symmetry}
\PopMa \PopF \crateMF = \PopFa \PopM \crateFM,
\end{equation}
from which we find that contact rates are symmetric in the sense that $\crateMF = \crateFM$ for all $a$, $b$, and similarly $\crateMM = \crateMMb$ and $\crateFF = \crateFFb$ for all $a < b$. A similar symmetry property does not hold for contact intensities.

Property~\eqref{contact-rate-symmetry} implies that data on age group $a$ informs contact rates in both age dimensions, which we will exploit heavily below. If the survey captures participants for all possible age groups, i.~e. $\AgesPartG = \mathcal{B}$, the estimation problem reduces to $B\times B + B \times (B+1)/2\times 2 = B(2B+1)$ free contact rate parameters rather than $4B^2$ free parameters, an almost 50\% reduction. To take advantage of these self-consistency constraints, we follow~\cite{van_de_kassteele_efficient_2017} and expand Eq~\eqref{lkl-contact-patterns-b} to
\begin{subequations}\label{random-functions}
\begin{align}
\log \cint & = \beta_0 + \bmf^{gh}(a,b) + \log(\Pop[b]{h}), \quad g=M, \: h=F, \: a, b\in\mathcal{B},\\
\log \cinthg & = \beta_0 + \bmf^{gh}(b,a) + \log(\Pop[b]{g}),\quad g=M, \: h=F, \: a, b\in\mathcal{B},\\
\log \cintgg & = \beta_0 + \bmf^{gg}(a,b) + \log(\Pop[b]{g}),\quad g\in \{M,F\}, \: a\leq b, \\
\log \cintgg & = \beta_0 + \bmf^{gg}(b,a) + \log(\Pop[b]{g}),\quad g\in \{M,F\}, \: a>b,
\end{align}
\end{subequations}
where $\beta_0 \in \mathbb{R}$ is a real-valued baseline parameter, and $\bmf^{MF}$, $\bmf^{MM}$, $\bmf^{FF}$ are three real-valued, random  functions of two-dimensional continuous inputs on the compact domain $[0,84]\times [0,84]$. Specifically, we model the $\bmf^{gh}$ through computationally efficient Gaussian process approximations as described below, and for ease of notation, simply write $\bmf^{gh}(a,b)$. The random functions act as age-age-specific offsets to the baseline parameter and thus capture the age structure in human contact intensities. Using random functions, we can estimate arbitrary age-specific contact patterns, which is important because human contact patterns have changed substantially since the COVID-19 pandemic with school closures and other non-pharmaceutical interventions. 

%
%
%
%
%
%

\subsection*{Recovering fine age structure from coarse data}
For COVIMOD and similar contact surveys, participants were asked to report their contacts in the coarse age groups
\begin{equation}
    \begin{split}
        c\in\mathcal{C} = \{& \text{0-4, 5-9, 10-14, 15-19, 20-24, 25-34, 35-44,} \\ &\text{45-54, 55-64, 65-69, 70-74, 75-79, 80-84} \}
    \end{split}
\end{equation}
to facilitate reporting because participants often do not know or remember the exact age of their contacts. Importantly, the exact age of the participant is known, and we can leverage this information through the symmetry property in  Eq~\eqref{contact-rate-symmetry} to estimate contact intensities at a much finer resolution. Because of this fundamental property, we call our resulting model the ``Bayesian rate consistency model". Using the shape-scale parameterisation in Eq~\eqref{lkl-contact-patterns}, it follows that
\begin{subequations}\label{coarse-lkl-contact-patterns}
\begin{align}
Y_{ac}^{gh} & = \sum_{b \in c} \ccnt \sim \text{NegBinomial}\left(\sum_{b \in c} \alpha_{ab}^{gh}, \frac{\odisp}{1 + \odisp}\right)\\
\eccnt & = \alpha_{ab}^{gh} \odisp\\
\log \eccnt & = \log \cint + \log \Part, \label{coarse-lkl-contact-patterns-b}
\end{align}
\end{subequations}
where $g,h\in\{M,F\}$, $a\in \AgesPartG$, $b\in \mathcal{B}$, and $c\in\mathcal{C}$. We will demonstrate below that the high-resolution contact intensities $\cint$ are identifiable from coarse contact data.

%
%
%
%
%
%

\subsection*{Estimating dynamics in contact patterns}
It is in principle, straightforward to extend the model (\ref{lkl-contact-patterns}-\ref{coarse-lkl-contact-patterns}) to capture time trends in contact patterns, but a particular challenge arises in the context of repeat surveillance. Across COVIMOD survey waves, many participants agreed to report data in multiple rounds, and analyses indicate that participants tend to repeat fewer contacts in subsequent surveys regardless of the time they were initially surveyed, a phenomenon we call reporting fatigue. The time trends in the primary data are thus confounded by longitudinal reporting behaviour. To control for reporting fatigue, we denote by $Y_{trac}^{gh}$ the number of contacts to individuals of age group $c$ and gender $h$ that are reported at survey time $t$ by all participants of age $a$ and gender $g$ who have participated $r$ time(s). All other notation extends analogously. 

In the simplest case, we introduce age-homogeneous reporting fatigue effects $\rho_r \in \mathbb{R}$ at repeat response times $r= 0,1,2,\dotsc$, and jointly model the longitudinal data with
\begin{subequations}\label{coarse-lkl-contact-dynamics}
\begin{align}
Y_{trac}^{gh} & \sim \text{NegBinomial}\left(\sum_{b \in c} \alpha_{trab}^{gh}, \frac{\odisp}{1 + \odisp}\right)\\
\mu_{trab}^{gh}  & = \alpha_{trab}^{gh} \odisp\\
\log \mu_{trab}^{gh} & = \log m_{tab}^{gh} + \rho_r + \log(\Parttr),
\end{align}
\end{subequations}
where $t=1,2,\dotsc$ indicates the survey waves, $r= 0,1,2,\dotsc$ repeat surveillance, and $\rho_0 = 0$ for ease of notation. Here, reporting fatigue is captured by negative $\rho_r$ (that decreases with $r$), which in turn will adjust the contact intensities $\log m_{tac}^{gh}$ in follow-up survey rounds to higher estimates than the primary data suggest. 
New participants entered the COVIMOD survey in each survey wave, so we have data to provide independent information on the contact dynamics and reporting fatigue, with the model borrowing strength across all the data available.

A second challenge is estimating the dynamics in contact patterns while accounting for missing \& aggregated contact reports. We denote the number of missing \& aggregate contact reports by participants of age $a$ and gender $g$ in survey wave $t$ by $T_{ta}^g$, and in addition, consider the number of total contacts with detailed age information by participants of age $a$ and gender $g$ in survey wave $t$ by $Y_{ta}^g = \sum_{r,c, h} Y_{trac}^{gh}$. Thus, we can calculate the proportion of contacts that are reported with detailed age information,
\begin{equation}
    S_{ta}^g =  Y_{ta}^g / (Y_{ta}^g + T_{ta}^g),
\end{equation}
and use it as an additional offset term in the linear predictor,
\begin{equation}\label{eq_group-contacts}
    \log \mu_{trab}^{gh} = \log m_{tab}^{gh} + \rho_r + \log(\Parttr) + \log(S_{ta}^g).
\end{equation}
In practice, if increasingly many participants only provide aggregated contact reports, we have that $S_{ta}^g < 1$, and in turn, this will adjust the contact intensities $\log m_{tab}^{gh}$ in later survey waves to higher estimates than the data with full age-specific details suggest. 

%
%
%
%
%
%

\subsection*{Non-parametric modelling of contact dynamics}
We regularise our inferences in high-dimensional parameter space by associating the random functions $\bmf_t^{MF}$, $\bmf_t^{MM}$, $\bmf_t^{FF}$ in Eq~\eqref{random-functions} with computationally efficient, zero-mean, two-dimensional Hilbert Space Gaussian Process approximation priors~\cite{xi_inferring_2022, riutort-mayol_practical_2022}. In the next sections, we will drop the time and gender sub- and superscripts to ease notation and present our modelling of a generic random function $\bmf$ that represents age structure in contact patterns.

Zero-mean two-dimensional Gaussian Processes (GPs) are powerful prior models for random functions. For any finite collection of two-dimensional inputs, the function values are multivariate normal with mean zero. We always have $AB$ count observations on the grid $x_1 = (a_1,b_1), \dotsc, x_{AB} = (a_A,b_B)$ defined by $A$ participant age groups in $\mathcal{A}$ and all possible $B$ population age groups in $\mathcal{B}$. The multivariate normal has then a covariance matrix $\bK\in\mathbb{R}^{AB\times AB}$ whose $i$, $j$\textsuperscript{th} entries are specified by a covariance kernel function $k(x_i, x_j)$. Here, we decompose the 2D kernel function for computational efficiency and model each component through squared exponential or Mat\'ern class kernels. Here, we have used squared exponential kernels as an example:
\begin{subequations}\label{squared-exp-kernels}
\begin{align}
& k\left( (a, b), (a', b') \right) = k^1(a, a')k^2(b, b') \label{kernel-kronecker-decomposition} \\
& k^1(a, a') = \mgnt_a^2\exp \left( -\frac{(a-a')^2}{2\lnsc_a^2} \right)\label{eq_kernel-eq}  \\
& k^2(b, b') = \mgnt_b^2\exp \left( -\frac{(b-b')^2}{2\lnsc_b^2} \right), 
\end{align}
\end{subequations}
where the scaling parameters $\mgnt_a$, $\mgnt_b$ control the magnitude of the random function in the corresponding dimension, and the lengthscale parameters $\lnsc_a$, $\lnsc_b$ control the bandwidth. The product in Eq~\eqref{kernel-kronecker-decomposition} is also known as Kronecker decomposition, because the covariance matrix $\bK$ equals the Kronecker product of the covariance matrices of the kernels with one-dimensional inputs, $\bK = \bK^2 \otimes \bK^1$, where the $i$, $j$th entry in $\bK^1\in\mathbb{R}^{A\times A}$ is given by $k^1(a_i, a_j)$ and the $i$, $j$th entry in $\bK^2\in\mathbb{R}^{B\times B}$ is given by $k^2(b_i, b_j)$. For computing purposes, we exploit that $\bK^1$, $\bK^2$ are positive semi-definite and decompose the covariance matrices as $\bK^1= \bL^1 {\bL^1}^\top$, $\bK^2= \bL^2 {\bL^2}^\top$, where the superscript $\top$ denotes transposition. Using the mixed product property of Kronecker operations, we  obtain 
\begin{equation}
    \bK = \left( \bL^2 \otimes \bL^1 \right)\left( \bL^2 \otimes \bL^1 \right)^{\top}.
\end{equation}
This shows that the zero-mean two-dimensional GP prior attached to the random function $\bmf$ on the $AB$ inputs $\bx=(x_1, \dotsc, x_{AB})$ can be obtained by linear transformation of $(AB)^2$ i.~i.~d. standard Gaussian random variables $\bz \sim \mathcal{N}(0,1)$, 
\begin{equation}\label{GP-representation}
    \bmf(\bx) = \left( \bL^2 \otimes \bL^1 \right) \bz = \text{vec}\left( \left( \bL^2 \Big(\bL^1 \: \text{reshape}(\bz, A, B) \Big)^{\top} \right)^{\top}\right).
\end{equation}
In Eq~\eqref{GP-representation}, the left-hand side denotes the $AB$-dimensional column vector of the random function evaluated at the inputs, and the right-hand side shows how the Kronecker product is calculated by a series of basic arithmetic operations. The $\text{reshape}$ operation transforms the $AB$ dimensional column vector $\bz$ column-wise into a $A\times B$ dimensional matrix, and the $\text{vec}$ operation flattens $A\times B$ dimensional matrices column-wise into an $AB$ dimensional column vector. 

Eq~\eqref{GP-representation} also shows that the computational cost of two-dimensional GPs is entirely determined by calculating, first, $\bL^1$, $\bL^2$ for each new set of GPs hyperparameters, and then, second, performing the arithmetic operations associated with $( \bL^2 \otimes \bL^1 )\bz$. We further use Hilbert Space Gaussian Process (HSGP) approximations\cite{solin_hilbert_2020} to each of the kernels $k^1$ and $k^2$ in Eq~\eqref{squared-exp-kernels} to substantially reduce the cost associated with the first step. For brevity, we refer readers to the excellent introductions to HSGPs in~\cite{xi_inferring_2022, riutort-mayol_practical_2022}, and here merely note that the stationary isotropic kernels can be expressed as an infinite sum that involves the spectral density $S$, eigenfunctions $\phi_i$ and eigenvalues $\lambda_i$, $i=1,\dotsc,\infty$, associated with a certain Laplacian eigenvalue problem on a compact domain $\Omega$ that is strictly larger than $\mathcal{B}$. For convenience, the input domain $\mathcal{B}$ is shifted with the midpoint at zero, and then $\Omega$ is written as $[-L, L]$ for some $L>0$. To ease notation, we continue to write the shifted inputs as $a_i$, and $b_i$ in what follows. The HSGP approximation $\HSGPk^1$ to $k^1$ on the domain $[-L^1,L^1]$ is then obtained by truncating the infinite sum to the first $M^1$ terms,
\begin{subequations}\label{hsgp-series}
\begin{align}
k^1(a, a') \approx \HSGPk^1(a, a') & = \sum_{j=1}^{M^1}S^1(\sqrt{\lambda^1_j})\phi^1_j(a)\phi^1_j(a'),\label{hsgp-series-sum}\\
S^1(\omega) & = \mgnt_a^2(2\pi\lnsc_a)\exp\left(- \lnsc_a^2\omega^2/2\right) \label{hsgp-series-S},\\
\sqrt{\lambda^1_j} &=  (j\pi)/(2L^1), \label{hsgp-series-ev}\\
\phi^1_j(x) &= \sqrt{1/L^1}\sin\left( \sqrt{\lambda^1_j}(x + L^1) \right).\label{hsgp-series-ef}
\end{align}
\end{subequations}
Note that the spectral density above is specific to squared exponential kernels and that expressions for Mat\'ern class kernels may be found in~\cite{riutort-mayol_practical_2022, carl_edward_rasmussen_gaussian_2006}. Crucially, the GP hyperparameters $\mgnt_a$, $\lnsc_a$ enter only in Eq~\eqref{hsgp-series-S}, and the eigenvalues and eigenfunctions are the same regardless of the GP hyperparameters and depend only on the domain boundary value $L^1$ together with the observed inputs $a$, $a'$. This speeds up Bayesian computations significantly because the eigenvalues in Eq~\eqref{hsgp-series-ev} and eigenfunctions in Eq~\eqref{hsgp-series-ef} can be precomputed once and for all. Rewriting Eq~\eqref{hsgp-series-sum} in matrix notation, we see that $\bL^1$ is approximated by
\begin{equation}\label{hsgp-L-approx}
\bL^1 \approx \tilde{\bL}^1 = \bm{\Phi}^1\sqrt{\Delta^1},
\end{equation}
where the $A\times M^1$ matrix $\bm{\Phi}^1$ has the $i,j$ entries $\phi^1_j(a_i)$, and the $M^1\times M^1$ matrix $\bm{\Delta}^1$ is diagonal with $j,j$ entries $S^1(\sqrt{\lambda^1_j})$. Again, $\bm{\Phi}^1$ does not depend on the GP hyperparameters and can be precomputed. The arithmetic operations in Eq~\eqref{hsgp-L-approx} can harness computationally efficient diagonal-post-multiply functions in many linear algebra libraries. The HSGP approximation to the $k^2$ kernel is analogous. The tuning parameters of the HSGP approximations are the integers $M^1$, $M^2$ and the boundary values $L^1$, $L^2$, and we determine these using established diagnostics~\cite{riutort-mayol_practical_2022}. The zero-mean Kronecker-decomposed HSGP prior associated with our random functions $\bmf$ on the input grid $\bx$ is then
\begin{equation}\label{HSGP-representation}
    \bmf(\bx) = \left( \tilde{\bL}^2 \otimes \tilde{\bL}^1 \right) \tilde{\bz} = \text{vec}\left( \left( \tilde{\bL}^2 \Big(\tilde{\bL}^1 \: \text{reshape}(\tilde{\bz}, M^1, M^2) \Big)^{\top} \right)^{\top}\right),
\end{equation}
where $\tilde{\bz}$ is a $M^1M^2$ dimensional column vector of i.~i.~d. standard normal random variables, and the non-negative hyperparameters are $\theta= (\mgnt_a,\lnsc_a,\mgnt_b,\lnsc_b)$.

\subsection*{Difference-in-age parameterisation}
Human contact patterns tend to concentrate among individuals of similar age and individuals with similar age gaps (parent-child, grandparent-child and grandparent-parent). To capture this diagonal structure in the simple Kronecker decomposed priors in Eq~\eqref{squared-exp-kernels} for our 2D random functions $\bmf$, we follow~\cite{vandendijck_cohort-based_2022} and define $\bmf$ on an  age by difference-in-age space rather than an age by age space. This amounts to rotating the age-by-age space by 45 degrees so that the peer-peer, parent-child, grandparent-child, and grandparent-parent contacts correspond to horizontal lines in the re-parameterised space and match the structure of our Kronecker decomposed priors~\eqref{squared-exp-kernels} (See Fig 1 of Vandendijck et al.~\cite{vandendijck_cohort-based_2022}). 

Specifically, we consider age differences $d\in\mathcal{D}=\{ -84, -83, \ldots, 83, 84 \}$, and re-parameterise the points $(a,b) \in \mathcal{A}\times\mathcal{B}$ to $(a,d) = d(a,b) = (a, b-a) \in \mathcal{A}\times\mathcal{D}$. The number of age differences $D = 169$ in $\mathcal{D}$ is larger than the number of one-year age groups $B = 85$, and we are only interested in the random functions evaluated on the original points, which we write as $\bmf(d(a,b))$ for all $(a,b)\in \mathcal{A}\times\mathcal{B}$. We will show below that the difference-in-age parameterisation is associated with higher estimation accuracy of typical diagonal and off-diagonal human contact patterns at higher computational costs that arise from the larger input space with an additional $A^2 - A$ elements.

\subsection*{Full Bayesian model and numerical inference}
To complete our model for inferring contact dynamics from longitudinal survey data, we specified commonly used priors on all remaining model parameters, leading for survey waves $t=1,\dotsc,5$, reporting repeats $r=0,\dotsc,4$, gender $g,h \in \{M,F\}$, participant age groups $a\in\mathcal{A}^{trg}$ and population age groups $b\in\mathcal{B}$ to
\begin{subequations} \label{eq:full-model}
\begin{align}
Y_{trac}^{gh} & \sim \text{NegBinomial}\left(\sum_{b \in c} \alpha_{trab}^{gh}, \frac{\odisp}{1+\odisp}\right)\\
\mu_{trab}^{gh}  & = \alpha_{trab}^{gh} \odisp\\
\log \mu_{trab}^{gh} & = \log m_{tab}^{gh} + \rho_r + \log(\Parttr) + \log(S^g_{ta})\\
\log m_{tab}^{gh} & = \beta_0 + \tau_t + \bmf_t^{gh}\big(d(a,b)\big) + \log(\Pop[b]{h}), \quad g=M, \: h=F, \: a, b\in\mathcal{B}\\
\log m_{tab}^{hg} & = \beta_0 + \tau_t + \bmf_t^{gh}\big(d(b,a)\big) + \log(\Pop[b]{g}),\quad g=M, \: h=F, \: a, b\in\mathcal{B}\\
\log m_{tab}^{gg} & = \beta_0 + \tau_t +\bmf_t^{gg}\big(d(a,b)\big) + \log(\Pop[b]{g}),\quad g\in \{M,F\}, \: a\leq b, \\
\log m_{tab}^{gg} & = \beta_0 + \tau_t + \bmf_t^{gg}\big(d(b,a)\big) + \log(\Pop[b]{g}),\quad g\in \{M,F\}, \: a>b,
\end{align}
\end{subequations}
and
\begin{subequations}
\begin{align}
\beta_0 & \sim \mathcal{N}(0, 10) \\
\rho_r & \sim \mathcal{N}(0, 1) \\
\tau_t & \sim \mathcal{N}(0, 1) \\
\odisp & \sim \text{Exponential}(1) \\
\bmf_{t}^{gh}\big(d(\bx)\big)|\mgnt_{ti}, \lnsc_{ti} & \sim \text{HSGP}(\bm{0}, \tilde{\bL}_t^{gh,2} \otimes \tilde{\bL}_t^{gh,1} ), \quad gh\in \{ MF, MM, FF\}, i = 1, 2 \\
\mgnt_{ti} & \sim \text{Cauchy}^+(0,1),\quad i = 1,2  \\
\lnsc_{ti} & \sim \text{InvGamma}(5, 5),\quad i = 1,2. 
\end{align}
\end{subequations}
Monte Carlo draws from the joint posterior distribution of all parameters were obtained with the probabilistic computing language \texttt{Stan}~\cite{carpenter_stan_2017} via the \texttt{cmdstanr} interface version 0.5.2. Eight chains were run in parallel for 500 warmup iterations and 1000 iterations thereafter. Initial sampling was facilitated by adding the nugget $10^{-13}$ to $\alpha_{trab}^{gh}$. We typically observed a small number of divergences in the NUTS algorithm, but these accounted for less than 0.005\% of samples and were considered to be of no concern. The typical minimum effective sample sizes were 1627, and the $\hat{R}$ convergence diagnostics were below $1.01$, indicating that the Markov chains converged and mixed well~\cite{vehtari_rank-normalization_2021, betancourt_conceptual_2018}. The corresponding trace plots are shown in the Supp.

\subsection*{Simulated social contact data}
To validate the Bayesian models, we created synthetic datasets that mimic the social contact patterns with some simplifications before the COVID-19 pandemic (pre-COVID-19) and during the pandemic (in-COVID-19). We also varied the participant sample size in our experiments to assess its effect on estimation accuracy. To reduce experiment run time, we limited participants and contacts ages 6 to 49 and assumed that contact intensity patterns do not vary by gender. We generated contact intensity patterns based on the crude estimated marginal contact intensities of targeted age groups, which we obtained from the POLYMOD (pre-COVID-19)~\cite{mossong_social_2008} and CoMix studies (in-COVID-19)~\cite{jarvis_quantifying_2020} studies. 
 
Contact intensities were set to be highest among individuals of similar age, mimicking age-assortative contact behaviour. To simulate parent-children contact dynamics, we define individuals between 6-18 as children and individuals between 30-39 as parents and increase the contact intensities between these two age categories. Similarly, individuals between 19-29 are defined as children of individuals from 40 to 49. The resulting patterns are shown in the top left panel of Fig~\ref{fig:sim-exp-pre} and Fig~\ref{fig:sim-exp-in}. For full details, we refer the readers to~\nameref{S2_Text}.

From the stylised contact intensity scenarios, we next randomly generated age- and gender-specific contact counts for five different participant size configurations, $N = 250, 500, 1000, 2000, 5000$, by sampling from a Poisson distribution such that $Y_{ab}^{gh} \sim \text{Poisson}(\lambda_{ab}^{gh})$ where $\lambda_{ab}^{gh} = \tilde{m}_{ab}^{gh}N_{a}^{g}$. We set $N_{a}^{g}$ such that the age-gender counts of the participants is representative of the 2011 German census population~\cite{statistische_amter_des_bundes_und_der_lander_zensus2011_2011}. To mimic the age reporting scheme in the COVIMOD surveys, we aggregate the simulated contact counts by $Y^{gh}_{ac} = \sum_{b \in c} Y_{ab}^{gh}$ where,
$$
c \in \mathcal{C}^{\text{sim}} = \{ \text{6-9, 10-14, 15-19, 20-24, 25-34, 35-44, 45-49}\},
$$
as illustrated in the top right panels of Fig~\ref{fig:sim-exp-pre}-\ref{fig:sim-exp-in}. In this fashion, we generated 10 replicate datasets for each experiment configuration (pre-COVID-19/in-COVID-19 and sample size) to obtain representative accuracy and runtime estimates. As we were interested in the performance of our GP-based models for the age by age and difference-in-age by age parameterisations, we ran cross-sectional versions of our model in Eq~\eqref{eq:full-model} where we dropped the time and repeat terms.

\section*{Results}

\subsection*{Contact patterns by 1-year age band can be estimated}

\begin{figure}[!t]
    \includegraphics[width=\textwidth]{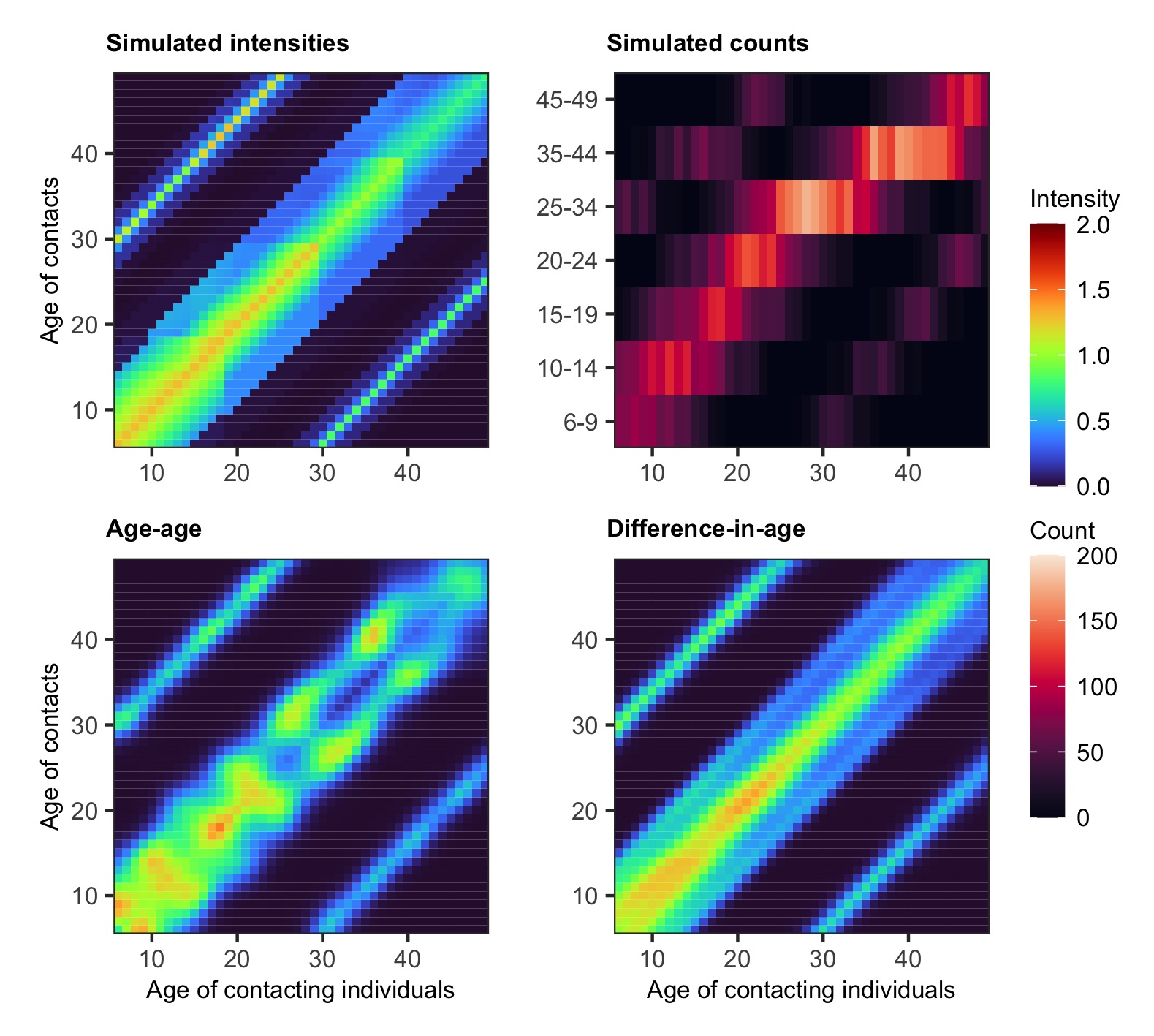}
    \caption{{\bf Pre-COVID19 scenario simulation experiments.} 
    (Top left) Simulated social contact intensities for Male-Male contacts. (Top right) Simulated social contact counts for Male-Male contacts with a COVIMOD-like age aggregation scheme. (Bottom left) Estimated social contact intensities from the age-age parameterisation HSGP model. (Bottom right) Estimated social contact intensities from the difference-in-age parameterisation HSGP model.}
    \label{fig:sim-exp-pre}
\end{figure}

\begin{figure}[!t]
    \includegraphics[width=\textwidth]{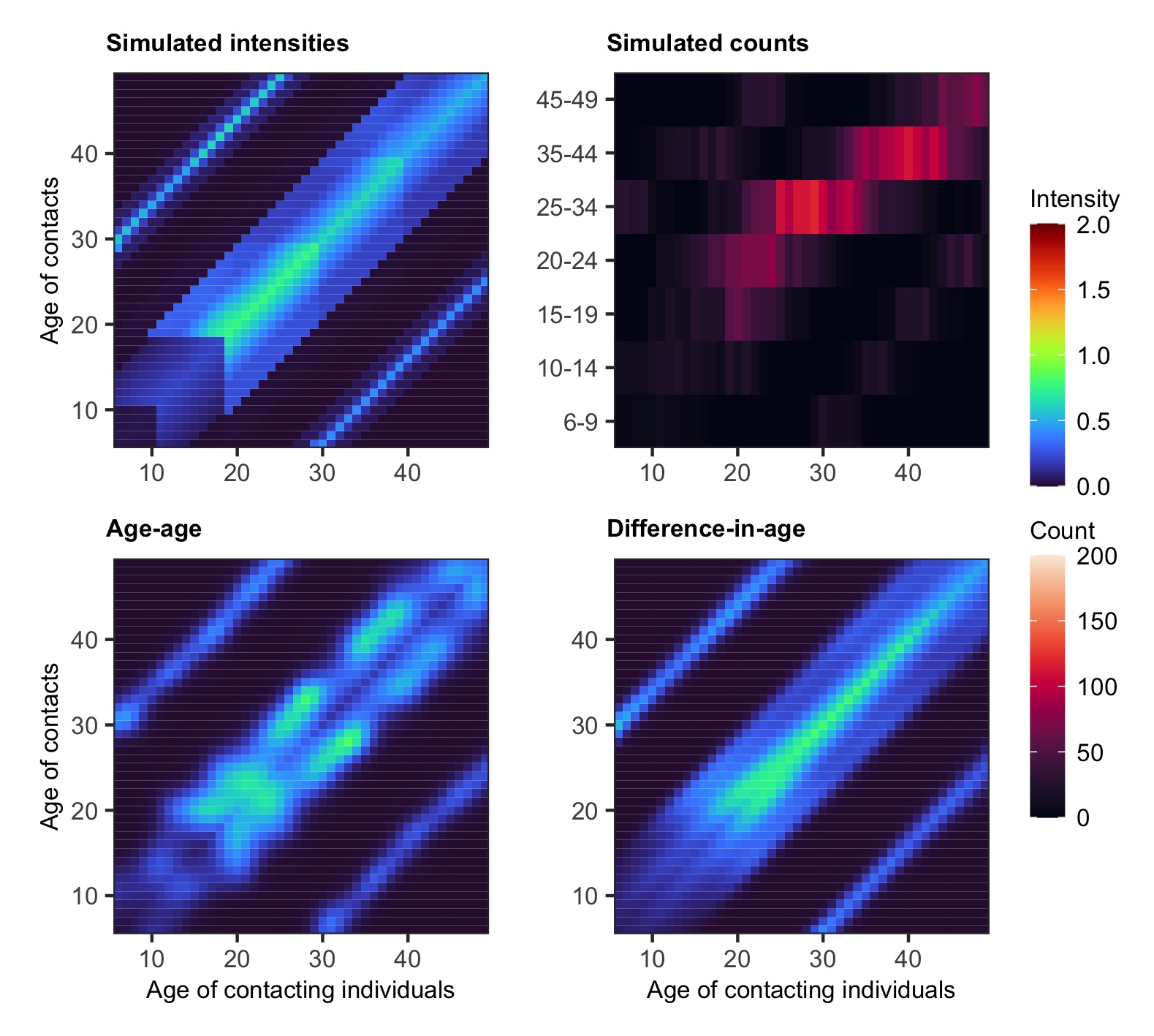}
    \caption{{\bf In-COVID19 scenario simulation experiments.} 
    (Top left) Simulated social contact intensities for Male-Male contacts. (Top right) Simulated social contact counts for Male-Male contacts with a COVIMOD-like age aggregation scheme. (Bottom left) Estimated social contact intensities from the age-age parameterisation HSGP model. (Bottom right) Estimated social contact intensities from the difference-in-age parameterisation HSGP model.}
    \label{fig:sim-exp-in}
\end{figure}

Fig~\ref{fig:sim-exp-pre} illustrates the fits of the Bayesian rate consistency model using the age-age and difference-in-age parameterisations for the pre-COVID-19 scenario, and Fig~\ref{fig:sim-exp-in} for the  in-COVID-19 scenario, both with a sample size of 2,000. The age-age parameterisation performed poorly for both simulation scenarios, especially in regions of the contact matrix where the degree of age aggregation is large, i.e., 10-year age intervals as opposed to 5-year age intervals. For such large reporting intervals of age groups, the contact intensity patterns that we could estimate with the age-age parameterisation showed idiosyncratic bimodal patterns along the main diagonal of the contact intensity matrices (bottom left panels in Fig~\ref{fig:sim-exp-pre}-\ref{fig:sim-exp-in}). In comparison, the difference-in-age parameterisation captured age-assortative contact patterns and the sub-diagonal parent-children contact patterns with much better accuracy. The estimated contact intensity patterns for other gender combinations and other simulation scenarios were qualitatively very similar and are reported in~\nameref{S3_Fig} and~\nameref{S4_Fig}.

\subsection*{Gaussian process approximations enable fast Bayesian inference}

In Table~\ref{table:sim-exp}, we numerically compare the performance of the Bayesian rate consistency model with various parameterisations for different sample sizes and scenarios in terms of estimation accuracy and computing runtimes. Specifically, to assess how well HSGP models can approximate full-rank 2DGP models, we ran simulations for both scenarios and parameterisations with a sample size fixed at 2000. We compared the model fits to the simulation truth in terms of the mean absolute error (MAE) based on the age-age-specific inferred contact intensities, expected log posterior density (ELPD), the percentage that the predicted values are inside the 95\% prediction intervals according to posterior predictive check (PPC) and the median running time.

\begin{table}[!t]
\begin{adjustwidth}{-2.25in}{0in}
\centering
\caption{
{\bf Comparison of performance on simulated data for different scenarios, models, sample sizes, and parameterisations.}}
\begin{tabular}{|c|c|c|c|c|c|c|c|}
\hline
\bf Scenario & \bf N & \bf Model & \bf Parameterisation & \bf MAE$^a$ & \bf ELPD$^b$ & \bf PPC$^c$ & \bf Runtime$^d$ \\ \thickhline
pre$^e$ & 2000 & 2DGP & age-age       & $9.17 \times 10^{-2}$ & -3224.1 & 99.9\% & 1.3 hours \\ \hline
pre & 2000 & HSGP & age-age           & $9.04 \times 10^{-2}$ & -3175.8 & 99.8\% & 0.5 hours \\ \hline
pre & 2000 & 2DGP & difference-in-age & $4.24 \times 10^{-2}$ & -3017.7 & 98.2\% & 27.5 hours\\ \hline
pre & 2000 & HSGP & difference-in-age & $4.44 \times 10^{-2}$ & -3027.6 & 98.5\% & 2.1 hours\\ \hline
in$^f$  & 2000 & 2DGP & age-age       & $4.91 \times 10^{-2}$ & -2745.4 & 99.9\% & 1.5 hours \\ \hline
in  & 2000 & HSGP & age-age           & $4.93 \times 10^{-2}$ & -2720.4 & 99.7\% & 0.4 hours \\ \hline
in  & 2000 & 2DGP & difference-in-age & $2.77 \times 10^{-2}$ & -2639.1 & 98.5\% & 16.7 hours\\ \hline
in  & 2000 & HSGP & difference-in-age & $2.85 \times 10^{-2}$ & -2674.5 & 98.7\% & 1.4 hours \\ \thickhline
pre & 5000 & HSGP & difference-in-age & $4.17 \times 10^{-2}$ & -3865.2 & 98.2\% & 2.7 hours \\ \hline
pre & 2000 & HSGP & difference-in-age & $4.44 \times 10^{-2}$ & -3027.6 & 98.5\% & 2.1 hours\\ \hline
pre & 1000 & HSGP & difference-in-age & $4.79 \times 10^{-2}$ & -2587.8 & 98.7\% & 1.5 hours \\ \hline
pre & 500  & HSGP & difference-in-age & $5.20 \times 10^{-2}$ & -2256.5 & 98.7\% & 1.7 hours \\ \hline
pre & 250  & HSGP & difference-in-age & $5.73 \times 10^{-2}$ & -1765.6 & 99.0\% & 0.7 hours \\ \hline
in  & 5000 & HSGP & difference-in-age & $2.60 \times 10^{-2}$ & -3357.8 & 98.4\% & 2.6 hours \\ \hline
in  & 2000 & HSGP & difference-in-age & $2.85 \times 10^{-2}$ & -2674.5 & 98.7\% & 1.4 hours \\ \hline
in  & 1000 & HSGP & difference-in-age & $3.23 \times 10^{-2}$ & -2245.7 & 99.0\% & 1.4 hours \\ \hline
in  & 500  & HSGP & difference-in-age & $3.27 \times 10^{-2}$ & -1716.6 & 99.1\% & 1.3 hours \\ \hline
in  & 250  & HSGP & difference-in-age & $4.94 \times 10^{-2}$ & -1327.2 & 99.5\% & 1.3 hours \\ \hline
\end{tabular}
\begin{flushleft} 
$^a$Mean absolute error, $^b$Expected log posterior density, $^c$Posterior predictive check, $^d$Median runtime, $^e$pre-COVID19 scenario, $^f$in-COVID19 scenario.
\end{flushleft}
\label{table:sim-exp}
\end{adjustwidth}
\end{table}

Next, the difference-in-age parameterisation achieved significantly better accuracy than the age-age parameterisation but also required more time to fit due to the introduction of additional $A^2 - A$ nuisance parameters. The computational toll of the difference-in-age parameterisation was most strongly reflected in the median runtimes under the full-rank 2DGPs, which could take more than a day to fit. Using HSGPs, we reduced median runtimes by more than 20-fold. 
However, the HSGP models resulted in a slight decrease in accuracy (MAE and ELPD) than full-rank 2DGP models.

Finally, the bottom section of Table~\ref{table:sim-exp} compares the accuracy of the difference-in-age HSGP models across survey sample sizes. In general, a larger sample size entailed longer computations. Smaller sample sizes led to less accurate estimates, with a 5-fold reduction in sample size resulting in approximately a 10-15\% increase in MAE for the pre-COVID-19 scenario and a 20-35\% increase in MAE for the in-COVID-19 scenario.

\subsection*{Modelling marked structures in age-specific contact patterns}
Social contacts are strongly structured by age, reflecting common behaviour and social norms around family size, reproductive age, schooling, and other factors ~\cite{vandendijck_cohort-based_2022}. In turn, smooth process kernels such as the squared exponential (Eq~\eqref{eq_kernel-eq}) may not be well suited to describe marked changes in contact intensities. On our simulated contact scenarios, we find indeed that Mat\'ern $\frac{5}{2}$ and Mat\'ern $\frac{3}{2}$ kernels performed better in comparison to squared exponential kernels in terms of accuracy (Table~\ref{table:sim-exp}, \nameref{S5_Table}). The difference in accuracy between Mat\'ern $\frac{3}{2}$ and Mat\'ern $\frac{5}{2}$ was small and qualitatively indistinguishable (\nameref{S3_Fig}, \nameref{S4_Fig}), and in the following results we considered the Mat\'ern $\frac{5}{2}$ kernel.

\subsection*{Model-based estimates of contact patterns in Germany by 1-year age groups}
\begin{figure}[!t]
\begin{adjustwidth}{-2.25in}{0in}
    \includegraphics[width=\linewidth]{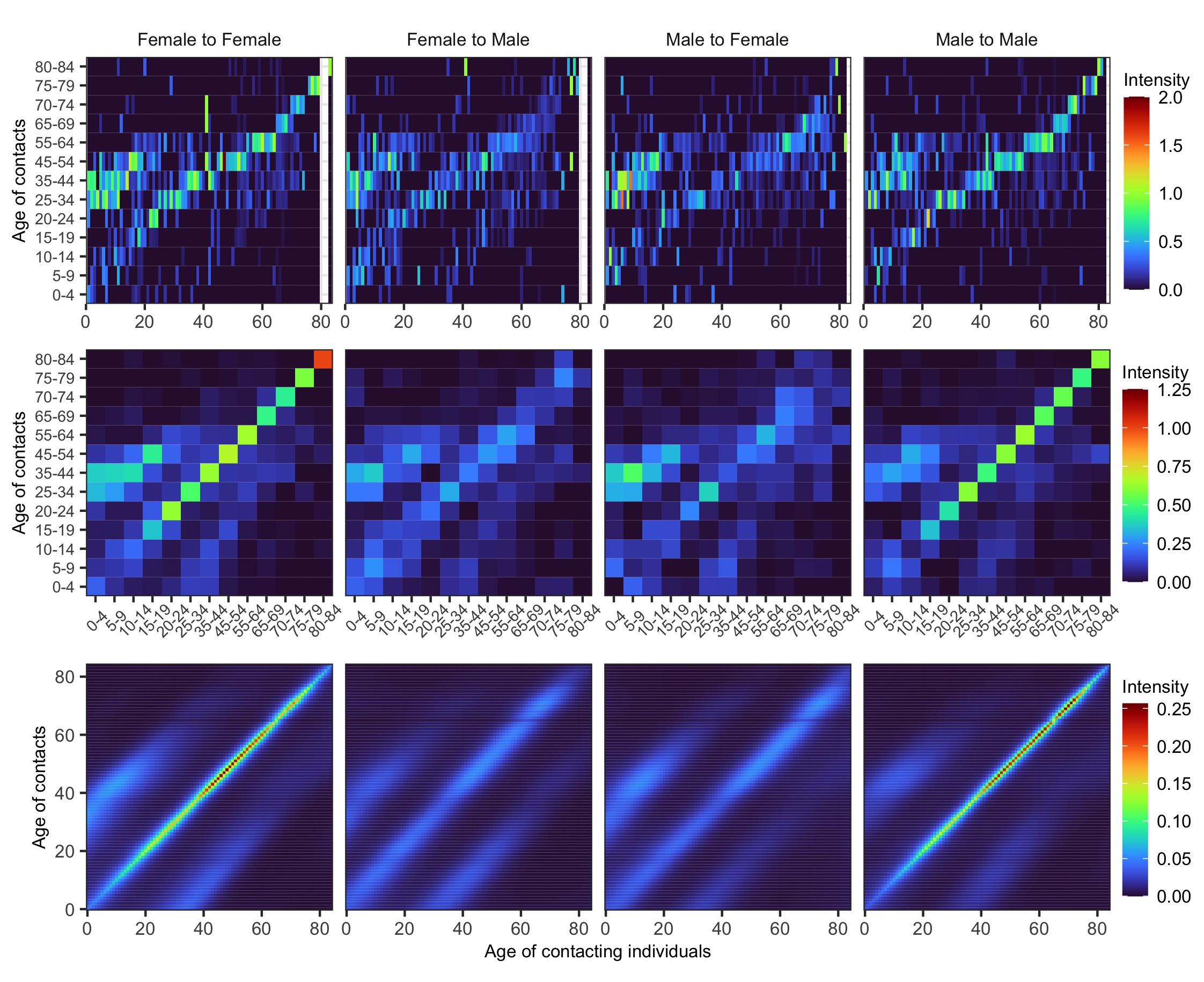}
    \begin{adjustwidth}{0.75in}{0in}
        \captionsetup{width=\linewidth}
        \caption{{\bf Empirical and estimated contact intensity patterns for COVIMOD wave 1}. (Top row) Crude empirical social contact intensity patterns, with crude contact intensities above a value of 3 truncated for visualisation purposes. There are some age groups with no participants, and they are represented by white vertical columns. (Middle row) Contact intensity patterns as estimated by the \texttt{socialmixr} R package~\cite{funk_socialmixr_2020}. (Bottom row) Contact intensity patterns are given by our Bayesian model. The exact runtime arguments for this comparison are given in script \texttt{figure-4.R} on our accompanying GitHub repository.}
        \label{fig:4}
    \end{adjustwidth}
\end{adjustwidth}
\end{figure}

The age- and gender-specific crude empirical contact intensities, contact intensity estimates from the \texttt{socialmixr} package~\cite{funk_socialmixr_2020} estimated via bootstrapping, and contact intensity estimates from the Bayesian rate consistency model for the first wave of COVIMOD are shown in Fig~\ref{fig:4}. Here, the ``crude" contact intensities were calculated from the data without any statistical modelling via
\begin{equation*}
    \hat{m}_{ac}^{gh} = \frac{Y_{ac}^{gh}}{N_a^g} \frac{1}{S_{a}^{g}},
\end{equation*}
where $Y_{ac}^{gh}$ denote the total number of contacts from participants of gender $g$ and age group $a$ to individuals of gender $h$ and age category $c$, $N_a^g$ are age- and gender-specific sample size, and $S_{a}^{g}$ are age- and gender-specific proportion of reports with complete age and gender information. The exact runtime arguments for this comparison are given in script \texttt{figure-4.R} on our accompanying GitHub repository. The crude estimates are sparse and fluctuate greatly, even between neighbouring age groups. They are also not symmetric in contact rates. In particular, some parents answered the survey on behalf of their children resulting in a larger asymmetry between parent-to-children versus children-to-parent contacts. 

The estimates from \texttt{socialmixr} aggregate the observations by the large age categories in which the contacts were reported and did not borrow available information through the exact age of participants to obtain higher resolution estimates. The \texttt{socialmixr} estimates are adjusted for symmetry in contact rates but not reporting fatigue or missing \& aggregate contact reports (see respectively Eq~\eqref{coarse-lkl-contact-dynamics} and Eq~\eqref{eq_group-contacts}). Furthermore, contacts with missing age information are imputed by sampling the missing age only from all contacts of the participants of the same age group~\cite{funk_socialmixr_2020}.

The estimated contact patterns from the Bayesian rate consistency model align with those estimated by \texttt{socialmixr} but provide much higher age resolution. Importantly, we achieve this higher resolution not via imputation but by logical constraints on who contacts whom in a closed population (recall Eq~\eqref{coarse-lkl-contact-patterns}). These constraints imply that data on the exact age of survey participants provides information on the exact contacts even though they are reported in coarse age brackets. The patterns reveal strong age-assortativeness in mixing patterns indicated by the high intensities on the main diagonals of the contact intensity matrices shown in the bottom row of Fig~\ref{fig:4}. Lying approximately 30 years away from the main diagonal, two strips of high contact intensity fade with increasing age and correspond to inter-generational contacts between parents and children. This age-dependent pattern persists over time (\nameref{S6_Fig} to \nameref{S9_Fig}). However, we show below that the increases in social contact intensities in subsequent waves were far from uniform.

\begin{figure}[!t]
\begin{adjustwidth}{-2.25in}{0in}
    \includegraphics[width=\linewidth]{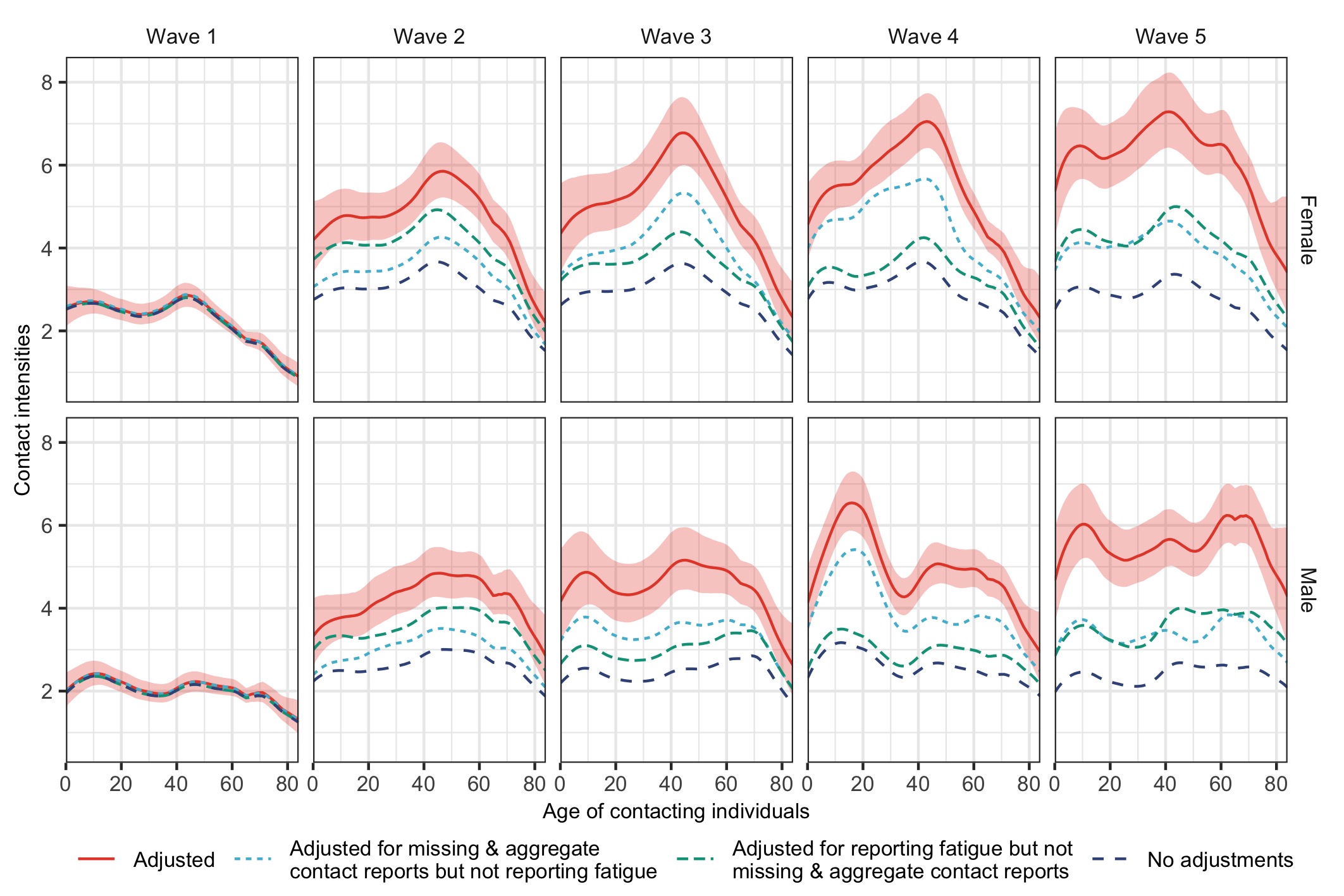}
    \begin{adjustwidth}{0.75in}{0in}
        \captionsetup{width=\linewidth}
        \caption{{\bf Marginal social contact intensities}. Dark blue dashed lines represent estimates without adjustments. Green dashed lines represent estimates adjusted for reporting fatigue but not missing \& aggregated contact reports. Turquoise dashed lines represent estimates adjusted for missing \& aggregated contact reports but not for reporting fatigue. Red solid lines represent estimates that are adjusted for both. Red bands represent 95\% credible intervals for adjusted estimates.}
        \label{fig:5}
    \end{adjustwidth}
\end{adjustwidth}
\end{figure}

\subsection*{Controlling for time-varying reporting effects}
We can sum the contacts' age dimension of estimated contact intensities to obtain the average number of contacts from one person of age $a$ per day. For brevity, we call these ``marginal" contact intensities. In Fig~\ref{fig:5}, we show the estimated marginal contact intensities under the Bayesian rate consistency model, which simultaneously accounts for the time-varying reporting effects that emerge through reporting fatigue (repeat participation in the longitudinal COVIMOD survey) and missing and aggregate contact reports (participants unable to list contacts individually), and which are clearly present in the data as shown in Fig~\ref{fig:1} and \nameref{S1_Fig}. We compare these (adjusted) marginal contact intensities in Fig~\ref{fig:5} to those obtained without adjusting for missing \& aggregate contact reports (but adjusting for reporting fatigue), those obtained without adjusting for reporting fatigue (but adjusting for missing and aggregate contact reports), and those obtained without any adjustments for time-varying, missing \& aggregate contact reports or reporting fatigue. Importantly, the reporting effect sizes can be estimated simultaneously and thus borrow strength across data from neighbouring age groups and are mathematically consistent with the symmetry constraints in contact rates in closed populations. The results clearly show that adjusting for missing and aggregate contact reports and reporting fatigue significantly increased the estimated marginal contact intensities in a non-trivial manner that depends on the contribution of repeat survey participants to each survey wave. As contact reduction measures were progressively eased between wave 1 to wave 5 of the COVIMOD survey, the adjusted estimates are more consistent with the timeline of non-pharmaceutical interventions in Germany.

\subsection*{Overall time trends in contact intensities from May to July 2020 in Germany}
\begin{figure}[!t]
\begin{adjustwidth}{-2.25in}{0in}
    \includegraphics[width=\linewidth]{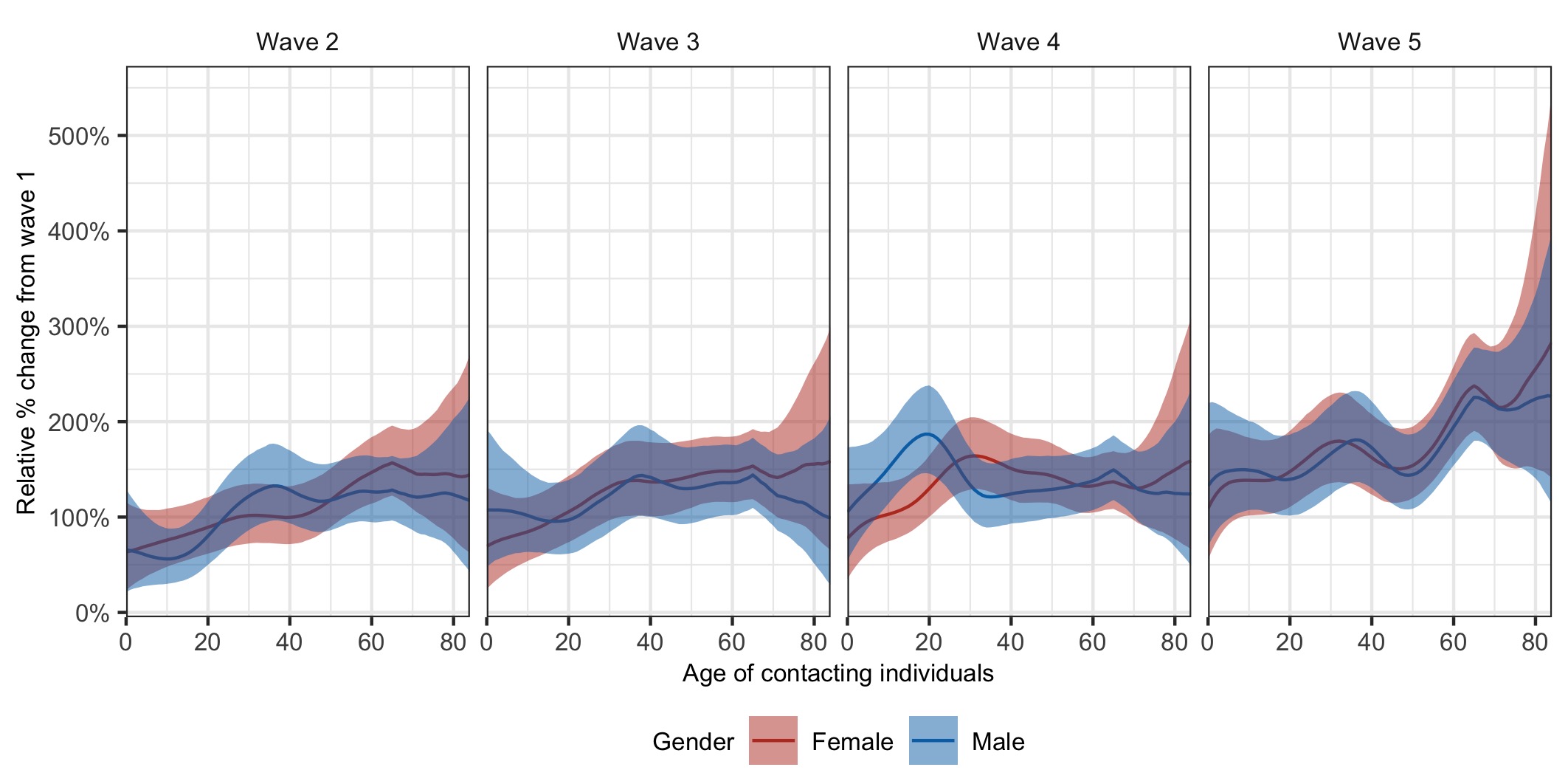}
    \begin{adjustwidth}{0.75in}{0in}
        \captionsetup{width=\linewidth}
        \caption{{\bf Relative percentage change of marginal contact intensities from wave 1}. The Red and blue lines represent posterior median estimates of the relative percentage change for adjusted marginal contact intensity estimates in females and males, respectively. Shaded ribbons represent 95\% credible intervals.}
        \label{fig:6}
    \end{adjustwidth}
\end{adjustwidth}
\end{figure}

Fig~\ref{fig:5} shows that, overall, the contact intensities increased consecutively from wave 1 to wave 5, although the increases were most substantial from wave 1 to wave 2. Fig~\ref{fig:6} illustrates the relative percentage increase in the marginal contact intensities relative to those in wave 1. Although there were marked differences in the contact patterns between men and women (Fig~\ref{fig:5}), the relative increases showed no significant differences between the two genders, which suggests that the gender differences in contact patterns may arise from underlying gender-dependent contact dynamics rather than non-pharmaceutical interventions. For waves 2 and 3, increases in contact intensities were higher in adults over 30 and were generally age-homogeneous. In wave 4, we observe a sharp increase in contacts among men approximately 20 years of age, but we find this pattern is sensitive to data pre-processing criteria as we explain below. In wave 5, the increase in contact intensities showed a rising pattern with age, where contact intensities increased most in older individuals.

\subsection*{Social contact intensities remained largely below pre-pandemic levels}

\begin{figure}[!t]
\begin{adjustwidth}{-2.25in}{0in}
    \includegraphics[width=\linewidth]{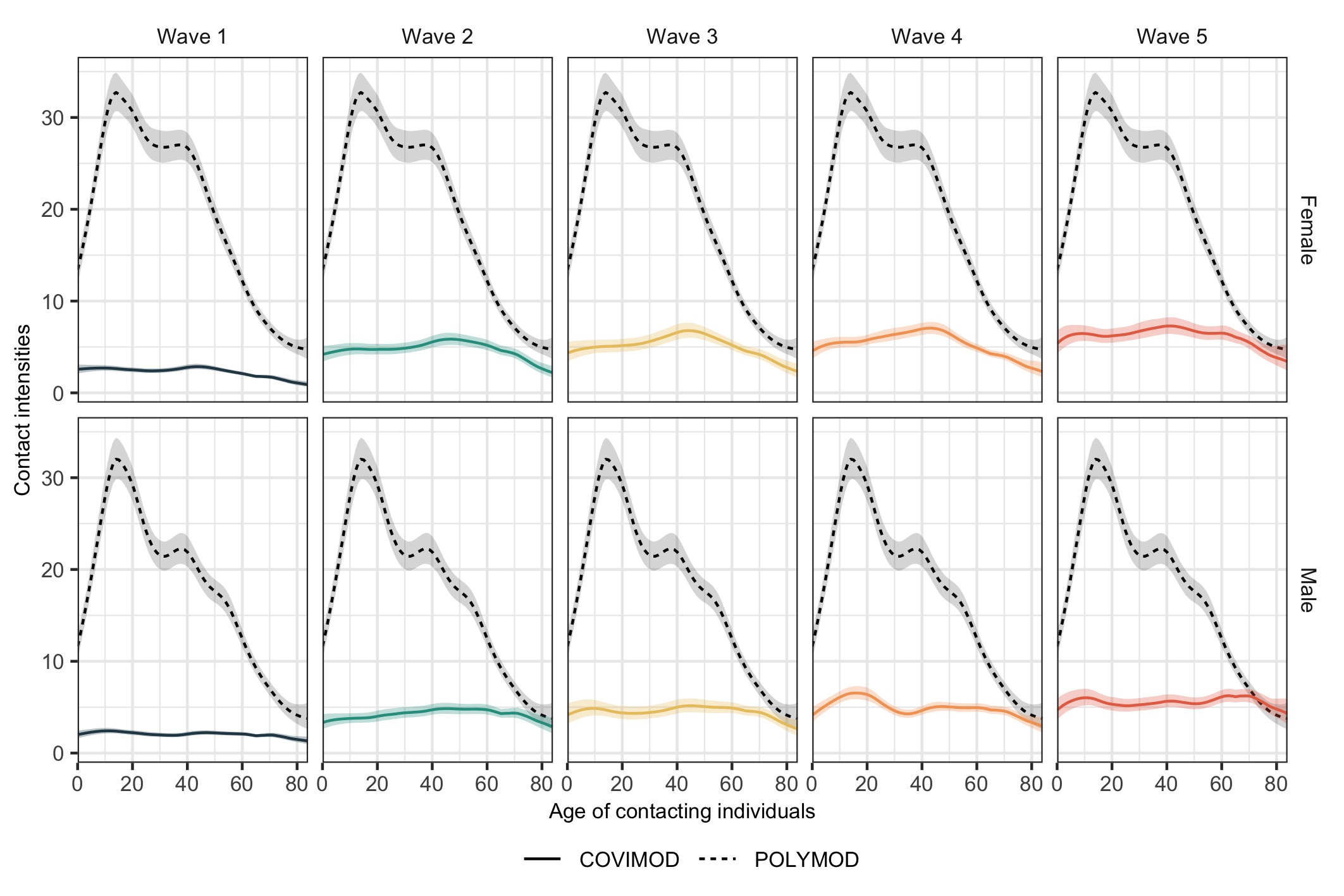}
    \begin{adjustwidth}{0.75in}{0in}
        \captionsetup{width=\linewidth}
        \caption{{\bf Marginal social contact intensities.} Solid lines represent COVIMOD estimates after adjusting for aggregate contacts and reporting fatigue. Dashed lines represent estimates from the pre-pandemic POLYMOD study. Shaded ribbons represent 95\% credible intervals.}
        \label{fig:7}
    \end{adjustwidth}
\end{adjustwidth}
\end{figure}

To gain more insights into the difference between social contact patterns before and during the COVID-19 pandemic, we compared the contact intensities seen during the first 5 waves of the COVIMOD study to those observed in the pre-pandemic POLYMOD study conducted between  2006 and 2008~\cite{mossong_social_2008} (Fig~\ref{fig:7}). We find that by wave 5, the contact intensities of individuals aged 70 were very similar to pre-pandemic levels. In contrast, the social contact intensities of all younger age groups in women and men remained substantially below pre-pandemic levels. Furthermore, similar to POLYMOD, we find that women between age 20 and 50 had more contacts than men of the same age range (\nameref{S10_Fig}).

\subsection*{Differential rebound of age-specific social contacts}

\begin{figure}[!t]
\begin{adjustwidth}{-2.25in}{0in}
    \includegraphics[width=\linewidth]{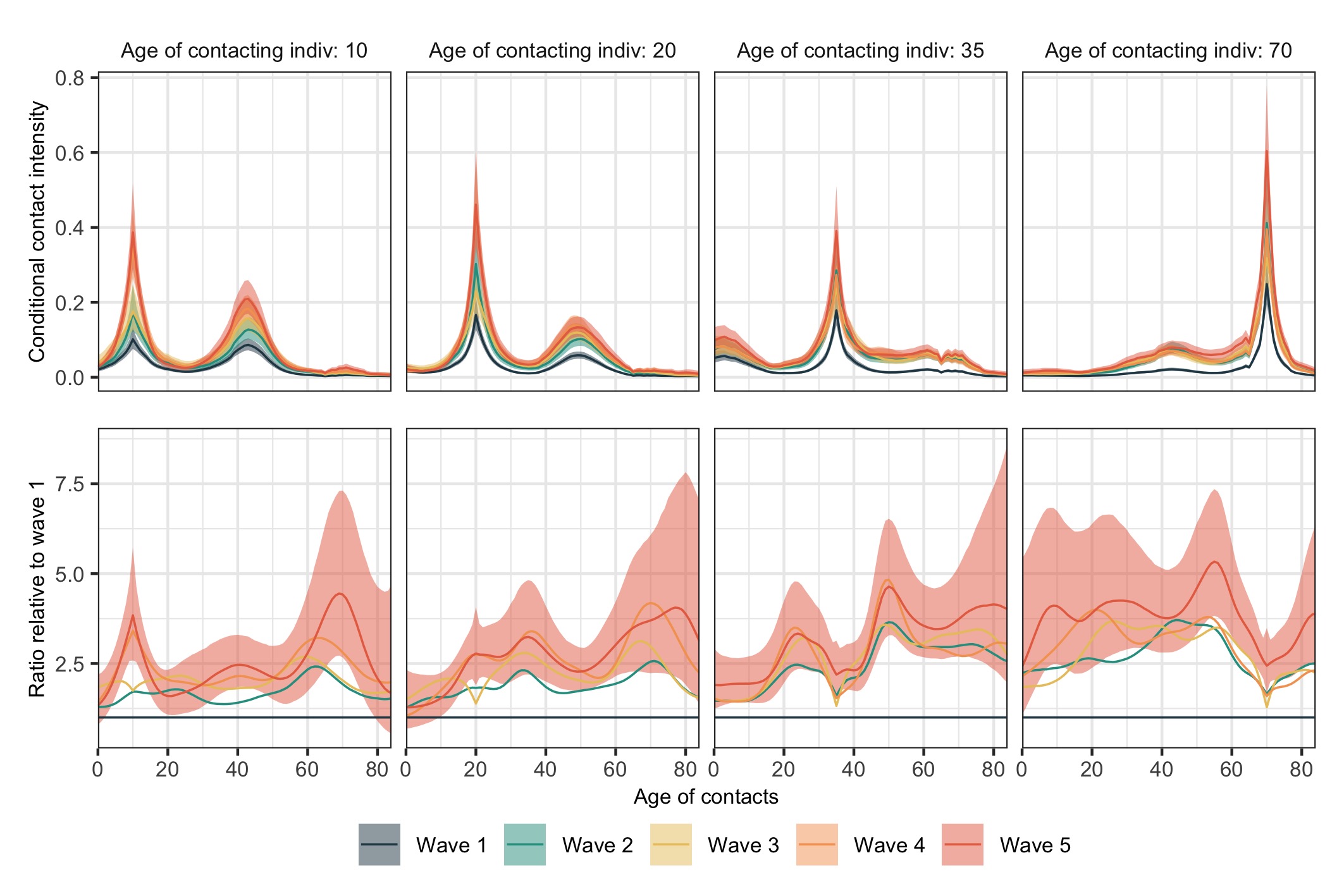}
    \begin{adjustwidth}{0.75in}{0in}
    \captionsetup{width=\linewidth}    
    \caption{{\bf Time evolution in age-specific social contact intensities.} The conditional contact intensities (top row) and relative change in the conditional contact intensities in waves 2 to 5 relative to those in wave 1 (bottom row) for individuals aged 10, 20, 35 and 70, respectively. Conditional contact intensities were aggregated across men and women. The colours represent different COVIMOD survey waves, and the shaded ribbons represent 95\% credible intervals. We only show credible intervals for wave 5 to reduce overlaps and ease interpretation.}
    \label{fig:8}
    \end{adjustwidth}
\end{adjustwidth}
\end{figure}

We next focus on characterising the dynamics in contact intensities for specific age groups. Intuitively, this corresponds to slicing the contact intensity matrices reported in Fig~\ref{fig:4}, \nameref{S6_Fig} to \nameref{S9_Fig} across rows for a fixed column, and for brevity, call these the ``conditional" contact intensities. In Fig~\ref{fig:8} (top), we illustrate the conditional contact intensities for individuals aged 10, 20, 35, and 70 years to represent the contact intensities of school children, young adults, the working population, and the ageing population. For participants aged 10 and 20, we observe two peaks: a larger sharp peak corresponding to contacts between peers and a shorter rounded peak with individuals approximately 45-50 years older, representing contacts with their parents.
For participants aged 35, we observe an additional third peak with individuals aged 60 to 70, predominantly corresponding to contacts with their parents. Participants aged 70 generally mixed with individuals of a similar age, with some contact between individuals aged approximately 40 but almost no contact with individuals under 20. These core patterns of social contacts are present across all survey waves.

Fig~\ref{fig:8} (bottom) shows the ratio of the conditional contact intensities in survey waves 2, 3, 4 and 5 relative to those in wave 1. Ratios above 1 thus indicate increases in social contacts of one individual with age shown in column facets to individuals with age indicated on the x-axis in Fig~\ref{fig:8}. We find that the increases in social contacts were not homogeneous by the age of contacted individuals. Focusing on wave 5 relative to wave 1, we find that for children aged 10, the conditional contact intensities in individuals of the same age and roughly aged 70 rose particularly strongly. For individuals aged 20, increases in their social contacts were more homogeneous, except for those with young children, which remained similar to those seen in wave 1. For individuals aged 35, increases in their social contacts were concentrated in slighter younger and all older individuals, but not their peers, reflecting that individuals in the 35-year age group retained social contact with their peers during intense non-pharmaceutical interventions in wave 1. Individuals aged 70 homogeneously increased their conditional contact intensities with younger individuals.

\section*{Discussion}
We developed the Bayesian rate consistency model in order to regain the ability to quantify and characterise social contact patterns at high age resolution from contemporary, longitudinal survey data.
The key contributions of our model-based approach to estimating the trends in human contact patterns are as follows. Our approach can accurately estimate high-resolution social contact patterns from data that aggregate the age of contacts into large age bands. We also provide a unified framework to adjust for the confounding effects due to the aggregated reporting of contacts and reporting fatigue. These advancements are particularly important for COVID-era studies from which fine-age contact reporting is unavailable and in which participants contributed to multiple survey waves~\cite{coletti_comix_2020, backer_impact_2021, verelst_socrates-comix_2021, gimma_changes_2022}. We draw from methodologies which serve as principal workhorses in spatial statistics to map the landscape of social contacts~\cite{diggle_model-based_2007, ton_spatial_2017} and incorporate a recently developed GP approximation technique to alleviate the computational bottleneck which often plagues such spatial models~\cite{solin_hilbert_2020, riutort-mayol_practical_2022}. For more complex and high-dimensional models, e.g. those involving more survey waves and other participant demographics, recent advances in approximating GP priors with variational autoencoders such as priorVAE\cite{semenova_priorvae_2022} and $\pi$VAE~\cite{mishra_pivae_2022} may be incorporated within our fully Bayesian framework.

In applying our model to the first five waves of the COVIMOD study, we gained insights into the age- and gender-specific social contact dynamics and their evolution over time at high age resolution. Despite a rebound in contact intensities after the lifting of contact reduction measures, we found that the estimates were still considerably lower than the pre-pandemic estimates obtained through POLYMOD (Fig~\ref{fig:6}), indicating a sustained behavioural change. These results are consistent with from the CoMix arms in England~\cite{gimma_changes_2022}, Belgium~\cite{coletti_comix_2020}, and the Netherlands~\cite{backer_impact_2021}. This suggests that the sustained shifts in contact patterns might apply more generally beyond the individual sampling frames of each study. Relative to pre-pandemic estimates, the largest contact reductions occurred among children and young adults (Fig~\ref{fig:6}), possibly due to school closures and the transition to remote work. Such results may indicate that non-pharmaceutical interventions successfully reduced risk in groups with the highest infection risk in pre-pandemic times.

In general, our results can aid epidemiologists and policymakers by providing a clearer picture of how COVID-19 and other infectious respiratory illnesses are propagated through the population. They can be used to parameterise infectious disease models to obtain more realistic estimates for epidemiological quantities such as the reproduction number $R$~\cite{van_de_kassteele_efficient_2017, wallinga_optimizing_2010}. Contact estimates from the first wave of COVIMOD may be of particular interest as they represent the patterns during a time of stay-at-home orders across Germany and subsequent relaxation of non-pharmaceutical interventions. As such, they could be used as template social contact patterns for modelling and forecasting future pandemics~\cite{flaxman_estimating_2020, monod_age_2021}, and pandemic preparedness building~\cite{wallinga_optimizing_2010}. This point is particularly important because the dynamics in social contact patterns were highly non-homogeneous.
Further, despite a continued increase in contact intensity, COVID-19 cases remained stable during the analysis period (Fig~\ref{fig:1}). This may be because contact counts remained much lower than pre-pandemic estimates despite the ease in non-pharmaceutical interventions~\cite{tomori_individual_2021}. Other protective measures such as face masks, hygiene regulations including surface disinfection, remote work, and warmer summer weather may also have contributed to keeping infections at bay~\cite{leung_transmissibility_2021}.

Our work is not without limitations. First, some issues arise from the sampling methodology of the COVIMOD survey. Participants were recruited from an online panel for market research with email invitations, and although quota sampling was performed, the final samples were not fully representative of the population. Previous work proposed using post-stratification weights to re-scale the data, but a sensitivity analysis did not reveal large differences between weighted and unweighted estimates~\cite{tomori_individual_2021}. Participants consisted only of those with internet access who possibly adhered more to social distancing rules as such a demographic is more likely to respond to health surveys. Participants received guidance only through the text within the questionnaires, which may have been misinterpreted, and participants may report more contacts in paper-based surveys than in an online survey~\cite{beutels_social_2006}. Additionally, we truncate aggregate contact reports at 60, but different thresholds may lead to slight changes in the inference results.
Secondly, the difference-in-age parameterisation may not be appropriate if social contacts do not follow the pattern where high intensities lie on the contact matrix's main diagonal and sub-diagonals. This is relevant when investigators wish to conduct analyses for various contexts, e.g., work and transport, where contact patterns may not depend on age and age difference. However, it is easy for investigators to revert to the classical age-age parameterisation if they deem it more appropriate. We provide template \texttt{Stan} model files in the accompanying GitHub repository. Third, our fully Bayesian modelling framework is currently limited to analysing approximately 10 longitudinal survey rounds. While the recently proposed Hilbert Space Gaussian Process priors enable fast Bayesian inferences on cross-sectional data~\cite{xi_inferring_2022, solin_hilbert_2020, riutort-mayol_practical_2022}, additional research is needed to scale up the approach to survey data from 30 waves or more. Fourth, our model requires participant age information to be reported by 1-year age bands. The exact age of participants is usually recorded without error but not necessarily made publicly available~\cite{verelst_socrates-comix_2021}, which limits the applicability of the model-based approach developed here.

\section*{Conclusion}

In summary, we propose a novel model-based Bayesian framework for high-resolution estimation of age- and gender-specific social contact patterns. We validate our model on simulated social contact data for different scenarios and demonstrate the effectiveness of our approach. In applying our model to the COVIMOD contact survey, we provide a detailed picture of how social contact patterns evolved during Germany's first wave of the COVID-19 pandemic. This work promises to aid the understanding of contact behaviour, more realistic parameterisations of infectious disease models, and a deeper understanding of how infectious respiratory diseases are propagated through populations.

\section*{Supporting information}

\paragraph*{S1 Fig.}
\label{S1_Fig}
{\bf The number of complete contact reports and aggregated contact reports by age, gender, and COVIMOD wave.} Pink bars represent aggregated contact reports, light blue bars represent non-household contacts, and dark blue bars represent household contacts. Aggregate contact reports were truncated at 60 (90\textsuperscript{th} percentile of aggregate contact reports) to remove the effects of extreme outliers.

\paragraph{S2 Text}
\label{S2_Text}
{\bf The construction of simulated social contact patterns.} A detailed description of how the contact intensity patterns used in the simulation experiments are generated.

\paragraph*{S3 Fig.}
\label{S3_Fig}
{\bf Simulation experiment results for the pre-COVID19 scenario with different covariance kernels.}. From top to bottom: results for the squared exponential kernel, results for the Mat\'ern $\frac{5}{2}$ kernel, and results for the Mat\'ern $\frac{3}{2}$ kernel. All experiments were run with HSGP using the difference-in-age parameterisation models with $M^1=40$ (Number eigenfunctions on the difference-in-age dimension) and $M^2=20$ (Number of eigenfunctions on the contacts' age dimension). The sample size was fixed at $N=2000$.

\paragraph{S4 Fig.}
\label{S4_Fig}
{\bf Simulation experiment results for the in-COVID19 scenario with different covariance kernels.}. From top to bottom: results for the squared exponential kernel, results for the Mat\'ern $\frac{5}{2}$ kernel, and results for the Mat\'ern $\frac{3}{2}$ kernel. All experiments were run with HSGP using the difference-in-age parameterisation models with $M^1=40$ (Number eigenfunctions on the difference-in-age dimension) and $M^2=20$ (Number of eigenfunctions on the contacts' age dimension). The sample size was fixed at $N=2000$.

\paragraph{S5 Table.}
\label{S5_Table}
{\bf Comparison of different covariance kernels and number of basis functions for HSGP models.} Results were obtained with models using the difference-in-age parameterisation. The sample size was fixed at $N=2000$ throughout. $M^1$: The number of HSGP basis functions on the difference-in-age dimension. $M^2$: The number of HSGP basis functions on the contacts' age dimension. $^a$Mean absolute error, $^b$Expected log posterior density, $^c$Posterior predictive check, $^d$Median runtime, $^e$pre-COVID19, $^f$in-COVID19 scenario.

\paragraph*{S6 Fig.}
\label{S6_Fig}
{\bf Empirical and estimated social contact intensity patterns for COVIMOD wave 2.} (Top row) Crude empirical social contact intensity patterns, with crude contact intensities above a value of 3 truncated for visualisation purposes. There are some age groups with no participants, and they are represented by white vertical columns. (Middle row) Contact intensity patterns as estimated by the \texttt{socialmixr} R package~\cite{funk_socialmixr_2020}. (Bottom row) Contact intensity patterns are given by our Bayesian model. The exact runtime arguments for this comparison are given in script \texttt{figure-6-9.R} on our accompanying GitHub repository.

\paragraph*{S7 Fig.}
\label{S7_Fig}
{\bf Empirical and estimated social contact intensity patterns for COVIMOD wave 3.} (Top row) Crude empirical social contact intensity patterns, with crude contact intensities above a value of 3 truncated for visualisation purposes. There are some age groups with no participants, and they are represented by white vertical columns. (Middle row) Contact intensity patterns as estimated by the \texttt{socialmixr} R package~\cite{funk_socialmixr_2020}. (Bottom row) Contact intensity patterns are given by our Bayesian model. The exact runtime arguments for this comparison are given in script \texttt{figure-6-9.R} on our accompanying GitHub repository.

\paragraph*{S8 Fig.}
\label{S8_Fig}
{\bf Empirical and estimated social contact intensity patterns for COVIMOD wave 4.} (Top row) Crude empirical social contact intensity patterns, with crude contact intensities above a value of 3 truncated for visualisation purposes. There are some age groups with no participants, and they are represented by white vertical columns. (Middle row) Contact intensity patterns as estimated by the \texttt{socialmixr} R package~\cite{funk_socialmixr_2020}. (Bottom row) Contact intensity patterns are given by our Bayesian model. The exact runtime arguments for this comparison are given in script \texttt{figure-6-9.R} on our accompanying GitHub repository.

\paragraph*{S9 Fig.}
\label{S9_Fig}
{\bf Empirical and estimated social contact intensity patterns for COVIMOD wave 5.} (Top row) Crude empirical social contact intensity patterns, with crude contact intensities above a value of 3 truncated for visualisation purposes. There are some age groups with no participants, and they are represented by white vertical columns. (Middle row) Contact intensity patterns as estimated by the \texttt{socialmixr} R package~\cite{funk_socialmixr_2020}. (Bottom row) Contact intensity patterns are given by our Bayesian model. The exact runtime arguments for this comparison are given in script \texttt{figure-6-9.R} on our accompanying GitHub repository.

\paragraph*{S10 Fig.}
\label{S10_Fig}
{\bf Ratio of female to male marginal contact intensities for POLYMOD and COVIMOD}. Lines represent posterior median estimates of the ratio of female to male marginal contact intensities, i.e., $m_a^F / m_a^M$. A ratio of 1 (dashed lines) indicates no difference in contact intensities between genders. Shaded ribbons represent 95\% credible intervals.

\section*{Acknowledgments}
YuC gratefully acknowledges funding from the Imperial President’s PhD Scholarship program, and the EPSRC Centre for Doctoral Training in Modern Statistics and Statistical Machine Learning at Imperial and Oxford (EP/S023151/1); OR from the Bill \& Melinda Gates Foundation (OPP1175094) and
the Medical Research Council (MR/V038109/1); MM from the EPSRC Centre for Doctoral Training in Modern Statistics and Statistical Machine Learning at Imperial and Oxford (EP/S023151/1) and the Bill \& Melinda Gates Foundation (OPP1175094); SB acknowledge support from the MRC Centre for Global Infectious Disease Analysis (MR/R015600/1), jointly funded by the UK Medical Research Council (MRC) and the UK Foreign, Commonwealth \& Development Office (FCDO), under the MRC/FCDO Concordat agreement, and also part of the EDCTP2 programme supported by the European Union. S.B. acknowledges support from the Novo Nordisk Foundation via The Novo Nordisk Young Investigator Award (NNF20OC0059309). S.B.\ acknowledges support from the Danish National Research Foundation via a chair position. S.B.\ acknowledges support from The Eric and Wendy Schmidt Fund For Strategic Innovation via the Schmidt Polymath Award (G-22-63345). S.B.\ acknowledges support from the  National Institute for Health Research (NIHR) via the Health Protection Research Unit in Modelling and Health Economics.
This work was further supported by the Imperial College Research Computing Service, DOI: 10.14469/hpc/2232. COVIMOD is funded by intramural funds of the Institute of Epidemiology and Social Medicine, University of M\"{u}nster, and of the Institute of Medical Epidemiology, Biometry and Informatics, Martin Luther University Halle-Wittenberg, as well as by funds provided by the Robert Koch Institute, Berlin, the Helmholtz-Gemeinschaft Deutscher Forschungszentren e.V. via the HZEpiAdHoc ``The Helmholtz Epidemiologic Response against the COVID-19 Pandemic” project, the Saxonian COVID-19 Research Consortium SaxoCOV (co-financed with tax funds on the basis of the budget passed by the Saxon state parliament), the Deutsche Forschungsgemeinschaft (DFG, German Research Foundation, via the project SpaceImpact project number 458526380) and the Federal Ministry of Education and Research (BMBF) via the projects Respinow (project number 031L0298F) and OptimAgent (project number 031L0299J) and as part of the Network University Medicine (NUM) via the egePan Unimed project (funding code: 01KX2021). The authors would like to thank Christopher Jarvis, Kevin Van Zandvoort, Amy Gimma, John Edmunds and the entire CoMix team for allowing the COVIMOD team to use an adapted version of the CoMix questionnaire for COVIMOD and for their great cooperation. The authors would also like to thank the team at IPSOS-Mori for their work on implementing the COVIMOD survey.

\nolinenumbers


%
%
%

\clearpage

\begin{figure}[!t]
\begin{adjustwidth}{-2.25in}{0in}
    \includegraphics[width=\linewidth]{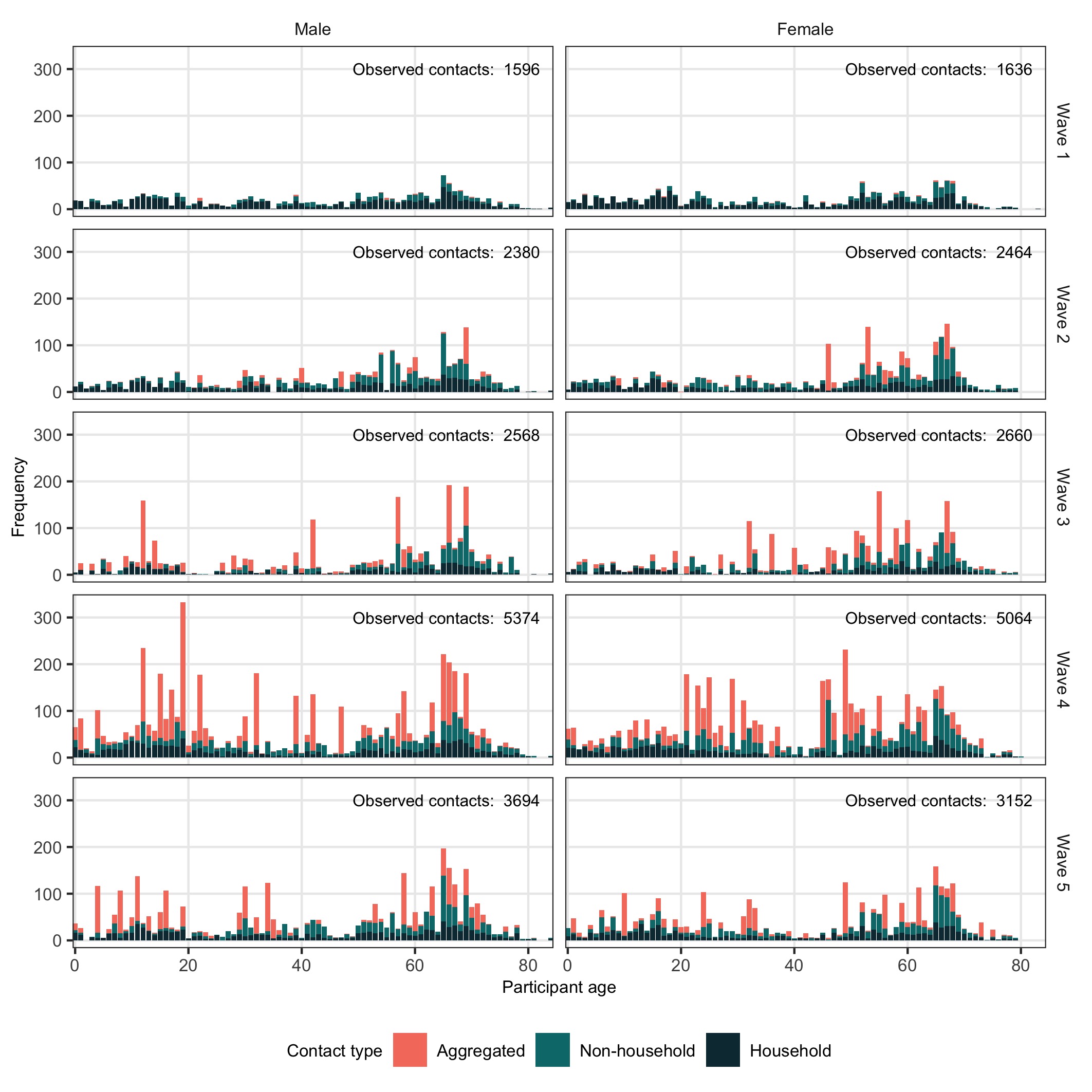}
    \begin{adjustwidth}{0.75in}{0in}
        \captionsetup{width=\linewidth}
        \caption*{{\bf S1 Fig. Distribution of observed contact counts by age, gender, wave, and type}. Pink bars represent contacts reported in aggregate, turquoise bars represent non-household contacts, and dark blue bars represent household contacts. Aggregate contact reports were truncated at 60 (90\textsuperscript{th}) percentile of aggregate contact reports to remove the effects of extreme outliers.}
    \end{adjustwidth}
\end{adjustwidth}
\end{figure}

\clearpage

\section*{S2 Text. The construction of simulated social contact patterns}

Existing social contact studies showed that human social contacts are highly dependent on age and age-difference~\cite{mossong_social_2008, van_de_kassteele_efficient_2017, jarvis_quantifying_2020, monod_age_2021, feehan_quantifying_2021}. Specifically, they reveal that contact intensities are strong between people of the same age and between parents and children. We aimed to mimic these real-world social contact dynamics in creating our simulation datasets. For both the pre-COVID-19 and in-COVID-19 scenarios, we limited the age range to 6 to 49 for simplicity and to reduce the run time of our experiments. We designate individuals in age bands 6-18, 19-29, 30-39 and 40-49 as children, young adults, adults, and middle-aged adults, respectively. A subset of people aged 30-39 and 40-49 was assumed to be the parents of children and young adults.
The following text describes the exact process by which the simulated contact intensities were generated. The R code may be found in script \texttt{R/sim-intensity-utility.R}, and the resulting contact intensity matrices can be found under \texttt{data/simulations/intensity} in the accompanying GitHub repository.

\subsection*{Pre-COVID-19 scenario}
Based on findings from POLYMOD study~\cite{mossong_social_2008},  we set the number of social contacts per day in children to roughly 32, which includes the 2 contacts with their parents. We assumed young adults have approximately 25 contacts per day with peers (i.e., people in the same age category) and 2 contacts with their parents. Adults were assumed to have approximately 20 daily contacts with peers and 4 with children. Middle-aged adults were assumed to have approximately 15 contacts per day with peers and 4 with young adults. We set contact intensities to be the highest for age-assortative contacts (i.e., between the same age groups) and decreased the intensity as the absolute age difference (AAD) increased. Additionally, we increased contact intensity for parent-child contacts (the AAD is 24) and decreased the intensity as the ADD increased. 
The following text describes the full data generation set-up for the pre-COVID-19 scenario. The contact intensity of age pairs not included in the list was set to 0.

\begin{enumerate}
    \item Participants aged 6-18: \\set the highest intensity as 2.5 ADD is zero, and decrease the intensity by 0.2 as the AAD increases to 8; \\
    set the intensity as 0.1 when AAD $\in[9:11]$; \\
    set the intensity as 0.03 when the AAD $\in[12:13]$;\\
    set the intensity as 0.01 when the AAD $\in[14:15]$;\\
set the second highest intensity as 0.8 for AAD is 24, and decrease the intensity by 0.5 as the AAD-24 increases to 1; \\
    set the intensity as 0.1 when the ADD-24 $\in[2:3]$;\\
    set the intensity as 0.01 when the AAD-24 $\in[4:5]$.
    
    \item Participants aged 19-29:\\ the highest intensity is 2.5 with 0.3 decrease by AAD until AAD = 5;\\
    set the intensity as 0.8 when AAD $\in[6:9]$; \\
    set the intensity as 0.04 when the AAD $\in[10:13]$;\\
    set the intensity as 0.01 when the AAD $\in[14:15]$;\\
set the second highest intensity as 0.8 for AAD is 24, and decrease the intensity by 0.5 as the AAD-24 increases to 1; \\
    set the intensity as 0.1 when the AAD-24 $\in[2:3]$;\\
    set the intensity as 0.01 when the AAD-24 $\in[4:5]$.
    
\item Participants aged 25-29, to generate the contacts with their children: \\set the intensity as 0.2 when the AAD-24 is from $\in[2:3]$;\\
set the intensity as 0.02 when the AAD-24 is from $\in[4:5]$; \\
set the intensity as 0.6 when the AAD-24 is 1 and participants aged 29.

\item Participants aged 30-39:\\ the highest intensity is 2 with 0.24 decrease by AAD until AAD = 5;\\
    set the intensity as 0.64 when AAD $\in[6:9]$; \\
    set the intensity as 0.03 when the AAD $\in[10:13]$;\\
    set the intensity as 0.01 when the AAD $\in[14:15]$;\\
set the second highest intensity as 1.6 for AAD is 24, and decrease the intensity by 1 as the AAD-24 increases to 1; \\
    set the intensity as 0.2 when the AAD-24 $\in[2:3]$;\\
    set the intensity as 0.02 when the AAD-24 $\in[4:5]$.
    
\item Participants aged 40-49:\\ the highest intensity is 1.5 with 0.18 decrease by AAD until AAD = 5;\\
    set the intensity as 0.5 when AAD $\in[6:9]$; \\
    set the intensity as 0.02 when the AAD $\in[10:13]$;\\
    set the intensity as 0.007 when the AAD $\in[14:15]$;\\
set the second highest intensity as 1.6 for AAD is 24, and decrease the intensity by 1 as the AAD-24 increases to 1; \\
    set the intensity as 0.2 when the AAD-24 $\in[2:3]$;\\
    set the intensity as 0.02 when the AAD-24 $\in[4:5]$.
\end{enumerate}

%
%
%
%

\subsection*{In-COVID-19 scenario}
Using the pre-COVID-19 scenario as a baseline, we incorporated the effects of contact reduction measures such as school closures and remote work for the in-COVID-19 scenario based on findings from the CoMix study~\cite{jarvis_quantifying_2020}. In children, the number of social contacts per day was set to roughly 3, among which 2 are with their parents. Young adults were assumed to have around 15 contacts per day with people of the same age and 2 with their parents. We assumed that adults have about 12 contacts with people of the same age and 2 with their children. Middle-aged adults were assumed to have approximately 10 contacts per day with peers and 2 with their children (young adults). Following the pre-COVID-19 scenario, contact intensities were set to be the highest for age-assortative contacts and decreased as the absolute age difference increased. Additionally, we increased contact intensity for parent-child contacts (the AAD is 24) and reduced the intensity as the ADD increased. 
The following text describes the full data generation setup for the in-COVID-19 scenario. The contact intensity of age pairs not included in the list was set to 0.

\begin{enumerate}
    \item Participants aged 6-10: \\set the highest intensity as 0.08 ADD is zero, and decrease the intensity by 0.007 as the AAD increases to 8; \\
    set the intensity as 0.003 when AAD $\in[9:11]$; \\
    set the intensity as 0.001 when the AAD is $\in[12:13]$;\\
    set the intensity as 3 $\times 10^{-4}$ when the AAD $\in[14:15]$;\\
set the second highest intensity as 0.8 for AAD is 24, and decrease the intensity by 0.5 as the AAD-24 increases to 1; \\
    set the intensity as 0.1 when the ADD-24 is $\in[2:3]$;\\
    set the intensity as 0.01 when the AAD-24 is $\in[4:5]$.

    \item Participants aged 11-18: \\set the highest intensity as 0.4 ADD is zero, and decrease the intensity by 0.035 as the AAD increases to 8; \\
    set the intensity as 0.015 when AAD $\in[9:11]$; \\
    set the intensity as 0.005 when the AAD $\in[12:13]$;\\
    set the intensity as 15 $\times 10^{-4}$ when the AAD $\in[14:15]$;\\
set the second highest intensity as 0.8 for AAD is 24, and decrease the intensity by 0.5 as the AAD-24 increases to 1; \\
    set the intensity as 0.1 when the ADD-24 $\in[2:3]$;\\
    set the intensity as 0.01 when the AAD-24 $\in[4:5]$.
    
    \item Participants aged 19-29:\\ the highest intensity is 1.5 with 0.18 decrease by AAD until AAD = 5;\\
    set the intensity as 0.48 when AAD $\in[6:9]$; \\
    set the intensity as 0.024 when the AAD $\in[10:13]$;\\
    set the intensity as 0.006 when the AAD $\in[14:15]$;\\
set the second highest intensity as 0.8 for AAD is 24, and decrease the intensity by 0.5 as the AAD-24 increases to 1; \\
    set the intensity as 0.1 when the AAD-24 $\in[2:3]$;\\
    set the intensity as 0.01 when the AAD-24 $\in[4:5]$.
    
\item Participants aged 25-29, to generate the contacts with their children: \\set the intensity as 0.1 when the AAD-24 is from $\in[2:3]$;\\
set the intensity as 0.01 when the AAD-24 is from $\in[4:5]$; \\
set the intensity as 0.6 when the AAD-24 is 1 and participants aged 29.

\item Participants aged 30-39:\\ the highest intensity is 1.25 with 0.15 decrease by AAD until AAD = 5;\\
    set the intensity as 0.4 when AAD $\in[6:9]$; \\
    set the intensity as 0.01875 when the AAD is $\in[10:13]$;\\
    set the intensity as 0.00625 when the AAD $\in[14:15]$;\\
set the second highest intensity as 0.8 for AAD is 24, and decrease the intensity by 0.5 as the AAD-24 increases to 1; \\
    set the intensity as 0.1 when the AAD-24 $\in[2:3]$;\\
    set the intensity as 0.01 when the AAD-24 $\in[4:5]$.
    
\item Participants aged 40-49:\\ the highest intensity is 1 with 0.12 decrease by AAD until AAD = 5;\\
    set the intensity as 0.33 when AAD $\in[6:9]$; \\
    set the intensity as 0.01375 when the AAD $\in[10:13]$;\\
    set the intensity as 0.00475 when the AAD $\in[14:15]$;\\
set the second highest intensity as 0.8 for AAD is 24, and decrease the intensity by 0.5 as the AAD-24 increases to 1; \\
    set the intensity as 0.1 when the AAD-24 $\in[2:3]$;\\
    set the intensity as 0.01 when the AAD-24 $\in[4:5]$.
\end{enumerate}

\begin{figure}[!t]
\begin{adjustwidth}{-2.25in}{0in}
    \includegraphics[width=\linewidth]{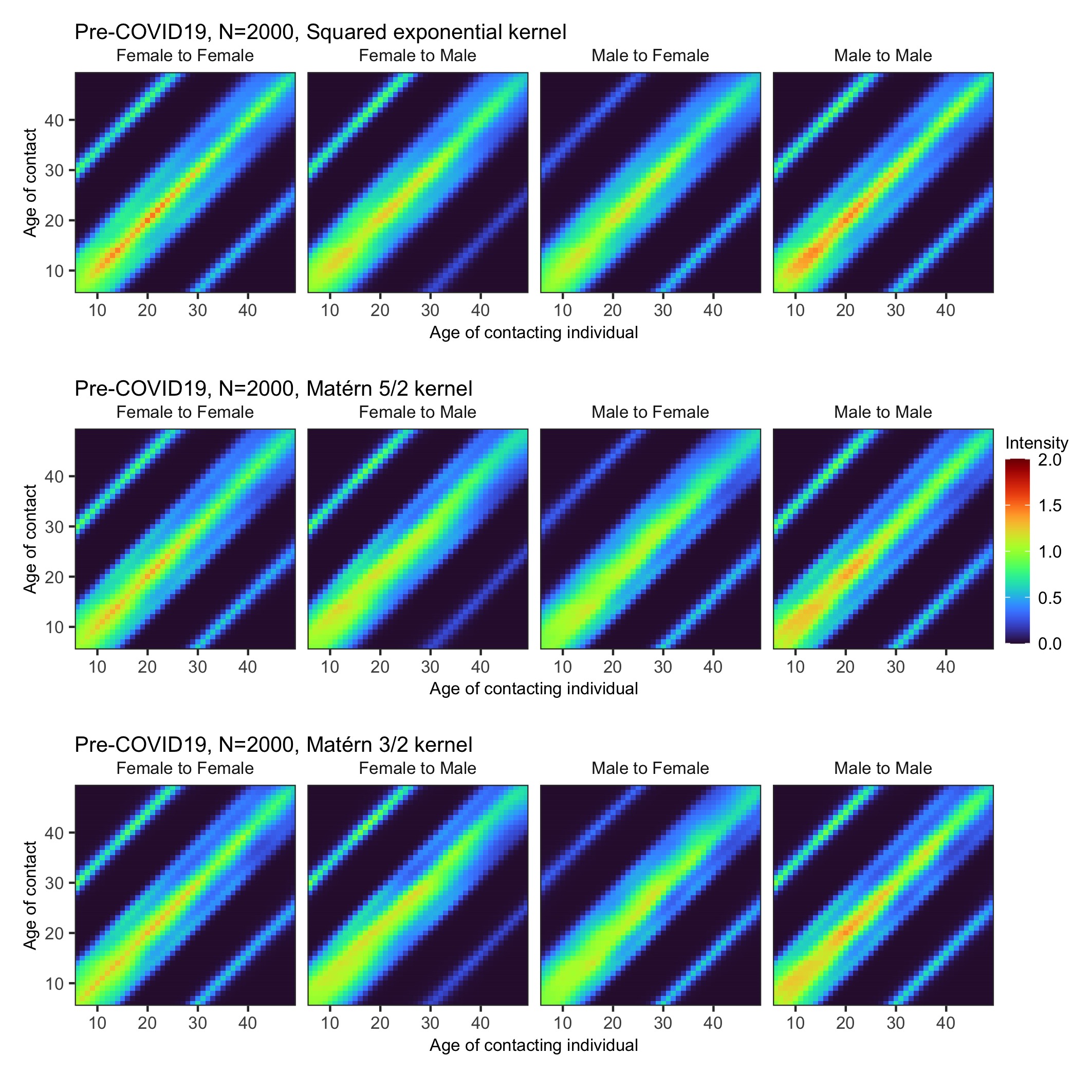}
    \begin{adjustwidth}{0.75in}{0in}
        \captionsetup{width=\linewidth}
        \caption*{{\bf S3 Fig. Simulation experiment results for the pre-COVID19 scenario with different covariance kernels.}. From top to bottom: results for the squared exponential kernel, results for the Mat\'ern $\frac{5}{2}$ kernel, and results for the Mat\'ern $\frac{3}{2}$ kernel. All experiments were run with HSGP using the difference-in-age parameterisation models with $M^1=40$ (Number eigenfunctions on the difference-in-age dimension) and $M^2=20$ (Number of eigenfunctions on the contacts' age dimension). The sample size was fixed at $N=2000$.}
    \end{adjustwidth}
\end{adjustwidth}
\end{figure}

\clearpage

\begin{figure}[!t]
\begin{adjustwidth}{-2.25in}{0in}
    \includegraphics[width=\linewidth]{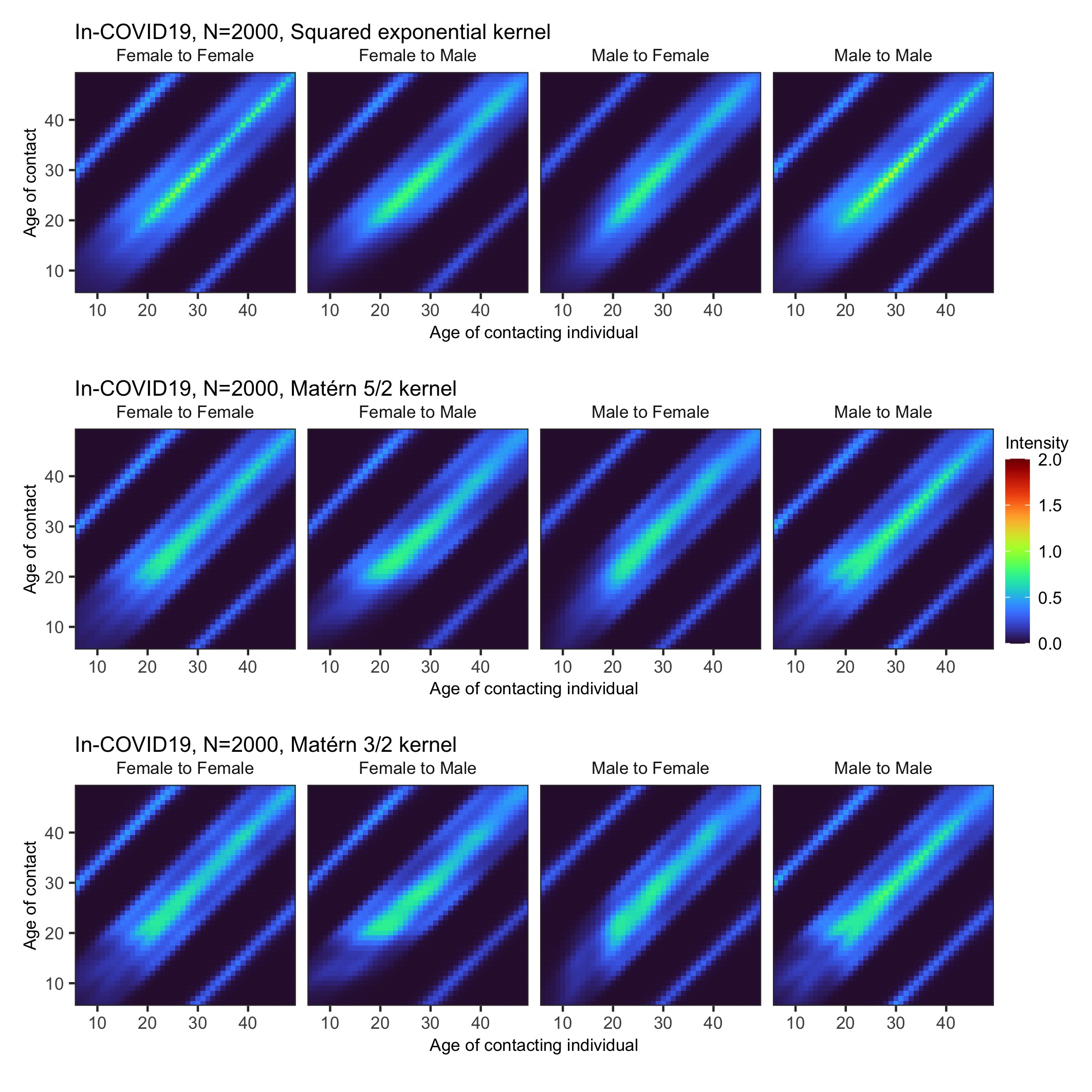}
    \begin{adjustwidth}{0.75in}{0in}
        \captionsetup{width=\linewidth}
        \caption*{{\bf S4 Fig. Simulation experiment results for the in-COVID19 scenario with different covariance kernels.}. From top to bottom: results for the squared exponential kernel, results for the Mat\'ern $\frac{5}{2}$ kernel, and results for the Mat\'ern $\frac{3}{2}$ kernel. All experiments were run with HSGP using the difference-in-age parameterisation models with $M^1=40$ (Number eigenfunctions on the difference-in-age dimension) and $M^2=20$ (Number of eigenfunctions on the contacts' age dimension). The sample size was fixed at $N=2000$.}
    \end{adjustwidth}
\end{adjustwidth}
\end{figure}

\clearpage

\begin{table}[!ht]
\begin{adjustwidth}{-2.25in}{0in} 
\centering
\caption*{
{\bf S5 Table. Comparison of performance on simulated data for different scenarios, covariance kernels, and number of basis functions.}}
\begin{tabular}{|c|c|c|c|c|c|c|c|}
\hline
\bf Scenario & \bf Kernel & \bf $M^1$ & \bf $M^2$ & \bf MAE$^a$ & \bf ELPD$^b$ & \bf PPC$^c$ & \bf Runtime$^d$ \\ \thickhline
pre$^e$ & SE           & 20 & 20 & $5.62 \times 10^{-2}$ & $-3120.4$ & 98.4\% & 1.6 hours \\ \hline
pre     & SE           & 30 & 20 & $5.00 \times 10^{-2}$ & $-3065.7$ & 98.3\% & 1.1 hours\\ \hline
pre     & SE           & 40 & 20 & $4.97 \times 10^{-2}$ & $-3064.2$ & 98.3\% & 1.2 hours\\ \hline
pre     & SE           & 40 & 30 & $4.97 \times 10^{-2}$ & $-3064.4$ & 98.3\% & 1.6 hours\\ \hline
pre     & Mat\'ern $5/2$ & 20 & 20 & $5.38 \times 10^{-2}$ & $-3099.4$ & 98.6\% & 1.4 hours \\ \hline
pre     & Mat\'ern $5/2$ & 30 & 20 & $4.79 \times 10^{-2}$ & $-3040.5$ & 98.5\% & 2.0 hours\\ \hline
pre     & Mat\'ern $5/2$ & 40 & 20 & $4.44 \times 10^{-2}$ & $-3027.6$ & 98.5\% & 2.1 hours\\ \hline
pre     & Mat\'ern $5/2$ & 40 & 30 & $4.41 \times 10^{-2}$ & $-3027.9$ & 98.5\% & 3.1 hours\\ \hline
pre     & Mat\'ern $3/2$ & 20 & 20 & $5.63 \times 10^{-2}$ & $-3094.6$ & 99.0\% & 2.1 hours \\ \hline
pre     & Mat\'ern $3/2$ & 30 & 20 & $4.81 \times 10^{-2}$ & $-3037.4$ & 98.8\% & 1.4 hours\\ \hline
pre     & Mat\'ern $3/2$ & 40 & 20 & $4.37 \times 10^{-2}$ & $-3024.2$ & 98.9\% & 1.2 hours\\ \hline
pre     & Mat\'ern $3/2$ & 40 & 30 & $4.36 \times 10^{-2}$ & $-3023.7$ & 98.9\% & 1.5 hours\\ \thickhline
in$^f$  & SE           & 20 & 20 & $3.73 \times 10^{-2}$ & $-2723.2$ & 98.3\% & 2.3 hours \\ \hline
in      & SE           & 30 & 20 & $3.27 \times 10^{-2}$ & $-2691.4$ & 98.4\% & 1.1 hours \\ \hline
in      & SE           & 40 & 20 & $3.27 \times 10^{-2}$ & $-2691.3$ & 98.4\% & 0.7 hours \\ \hline
in      & SE           & 40 & 30 & $3.27 \times 10^{-2}$ & $-2691.3$ & 98.4\% & 0.9 hours \\ \hline
in      & Mat\'ern $5/2$ & 20 & 20 & $3.56 \times 10^{-2}$ & $-2675.7$ & 98.6\% & 1.3 hours \\ \hline
in      & Mat\'ern $5/2$ & 30 & 20 & $2.97 \times 10^{-2}$ & $-2639.4$ & 98.6\% & 1.1 hours \\ \hline
in      & Mat\'ern $5/2$ & 40 & 20 & $2.85 \times 10^{-2}$ & $-2674.5$ & 98.7\% & 1.4 hours \\ \hline
in      & Mat\'ern $5/2$ & 40 & 30 & $2.85 \times 10^{-2}$ & $-2635.2$ & 98.7\% & 1.8 hours \\ \hline
in      & Mat\'ern $3/2$ & 20 & 20 & $3.41 \times 10^{-2}$ & $-2652.0$ & 98.9\% & 2.7 hours \\ \hline
in      & Mat\'ern $3/2$ & 30 & 20 & $2.86 \times 10^{-2}$ & $-2619.5$ & 99.0\% & 1.2 hours \\ \hline
in      & Mat\'ern $3/2$ & 40 & 20 & $2.70 \times 10^{-2}$ & $-2614.1$ & 99.0\% & 1.4 hours \\ \hline
in      & Mat\'ern $3/2$ & 40 & 30 & $2.69 \times 10^{-2}$ & $-2612.6$ & 99.0\% & 1.6 hours \\ \hline
\end{tabular}
\begin{flushleft} 
Results were obtained with models using the difference-in-age parameterisation. The sample size was fixed at $N=2000$ throughout. $M^1$: The number of HSGP basis functions on the difference-in-age dimension. $M^2$: The number of HSGP basis functions on the contacts' age dimension. $^a$Mean absolute error, $^b$Expected log posterior density, $^c$Posterior predictive check, $^d$Median runtime, $^e$pre-COVID19, $^f$in-COVID19 scenario.
\end{flushleft}
\end{adjustwidth}
\end{table}

\clearpage

\begin{figure}[!t]
\begin{adjustwidth}{-2.25in}{0in}
    \includegraphics[width=\linewidth]{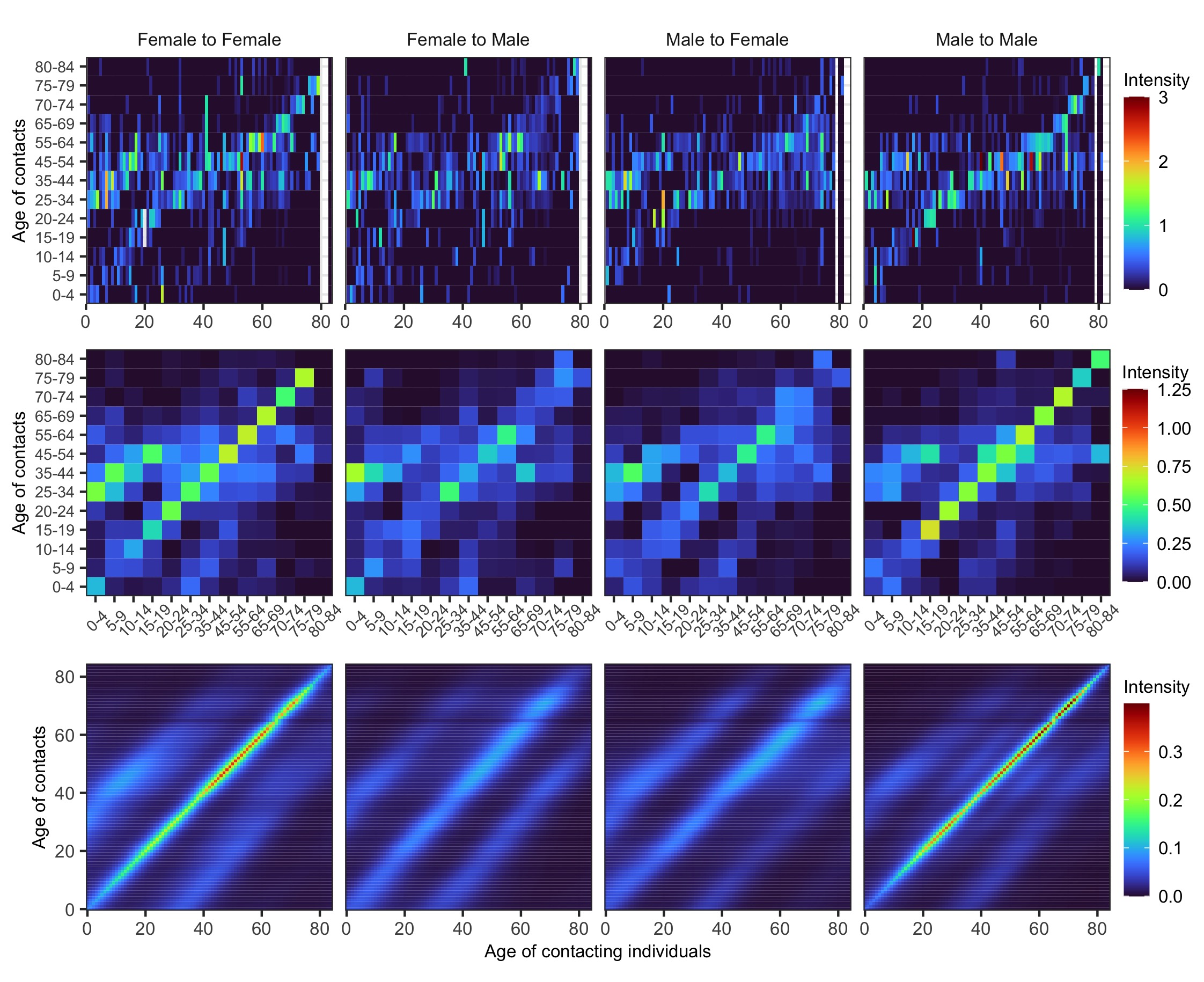}
    \begin{adjustwidth}{0.75in}{0in}
        \captionsetup{width=\linewidth}
        \caption*{{\bf S6 Fig. Empirical and estimated contact intensity patterns for COVIMOD wave 2}. (Top row) Crude empirical social contact intensity patterns, with crude contact intensities above a value of 3 truncated for visualisation purposes. There are some age groups with no participants, and they are represented by white vertical columns. (Middle row) Contact intensity patterns as estimated by the \texttt{socialmixr} R package~\cite{funk_socialmixr_2020}. (Bottom row) Contact intensity patterns are given by our Bayesian model. The exact runtime arguments for this comparison are given in script \texttt{figure-6-9.R} on our accompanying GitHub repository.}
    \end{adjustwidth}
\end{adjustwidth}
\end{figure}

\clearpage

\begin{figure}[!t]
\begin{adjustwidth}{-2.25in}{0in}
    \includegraphics[width=\linewidth]{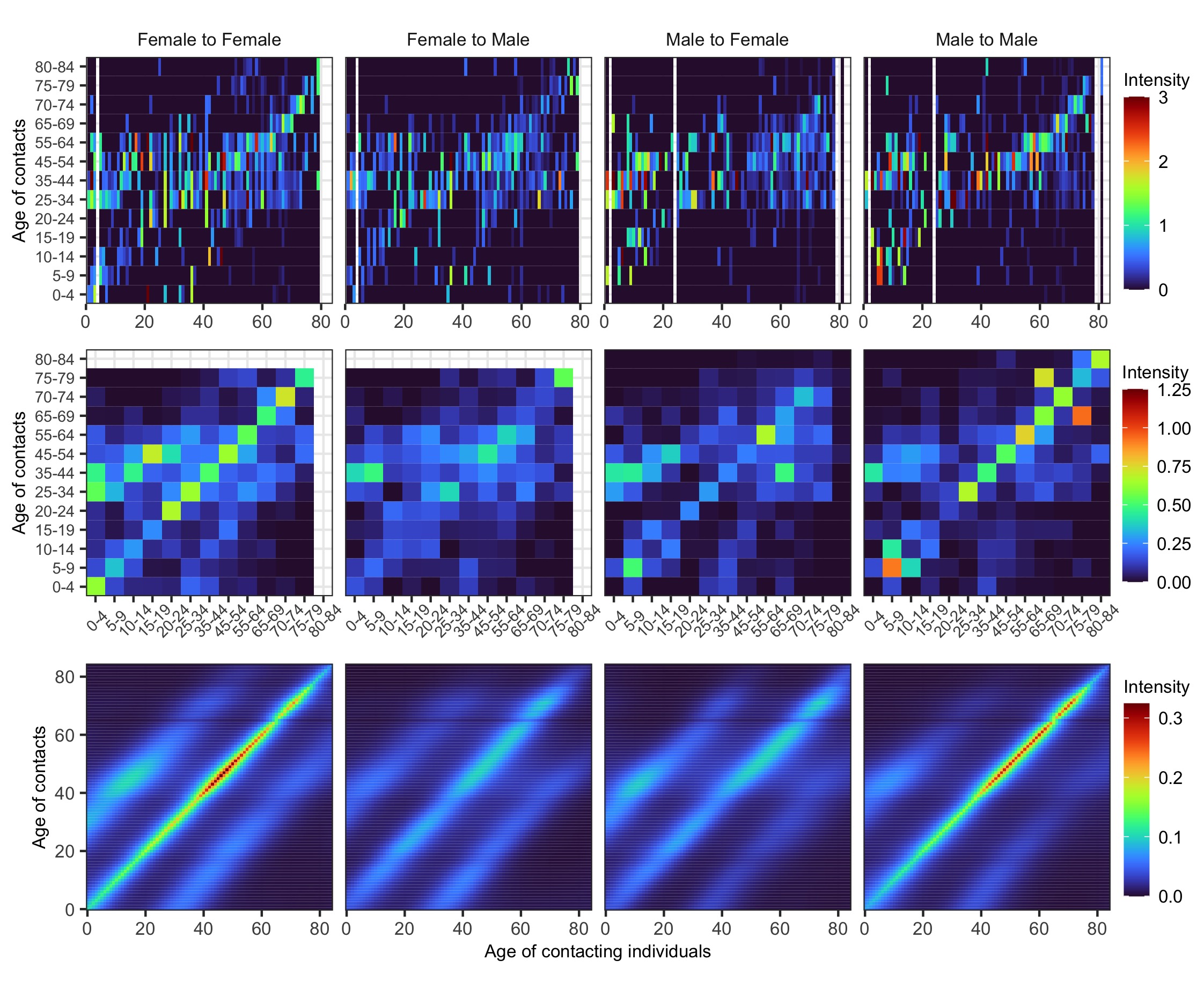}
    \begin{adjustwidth}{0.75in}{0in}
        \captionsetup{width=\linewidth}
        \caption*{{\bf S7 Fig. Empirical and estimated contact intensity patterns for COVIMOD wave 3}. (Top row) Crude empirical social contact intensity patterns, with crude contact intensities above a value of 3 truncated for visualisation purposes. There are some age groups with no participants, and they are represented by white vertical columns. (Middle row) Contact intensity patterns as estimated by the \texttt{socialmixr} R package~\cite{funk_socialmixr_2020}. (Bottom row) Contact intensity patterns are given by our Bayesian model. The exact runtime arguments for this comparison are given in script \texttt{sup-figure-6-9.R} on our accompanying GitHub repository.}
    \end{adjustwidth}
\end{adjustwidth}
\end{figure}

\clearpage

\begin{figure}[!t]
\begin{adjustwidth}{-2.25in}{0in}
    \includegraphics[width=\linewidth]{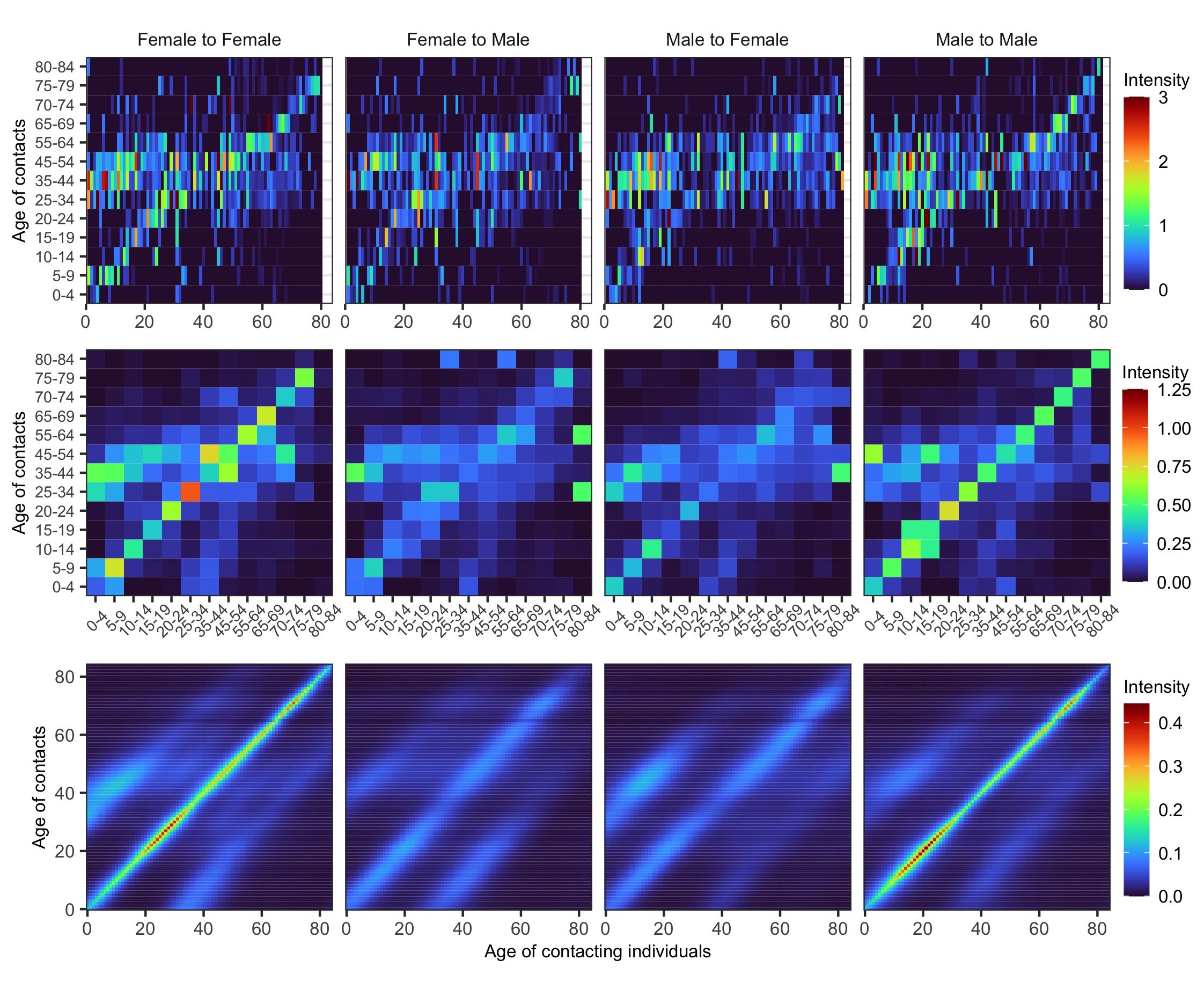}
    \begin{adjustwidth}{0.75in}{0in}
        \captionsetup{width=\linewidth}
        \caption*{{\bf S8 Fig. Empirical and estimated contact intensity patterns for COVIMOD wave 4}. (Top row) Crude empirical social contact intensity patterns, with crude contact intensities above a value of 3 truncated for visualisation purposes. There are some age groups with no participants, and they are represented by white vertical columns. (Middle row) Contact intensity patterns as estimated by the \texttt{socialmixr} R package~\cite{funk_socialmixr_2020}. (Bottom row) Contact intensity patterns are given by our Bayesian model. The exact runtime arguments for this comparison are given in script \texttt{figure-6-9.R} on our accompanying GitHub repository.}
    \end{adjustwidth}
\end{adjustwidth}
\end{figure}

\clearpage

\begin{figure}[!t]
\begin{adjustwidth}{-2.25in}{0in}
    \includegraphics[width=\linewidth]{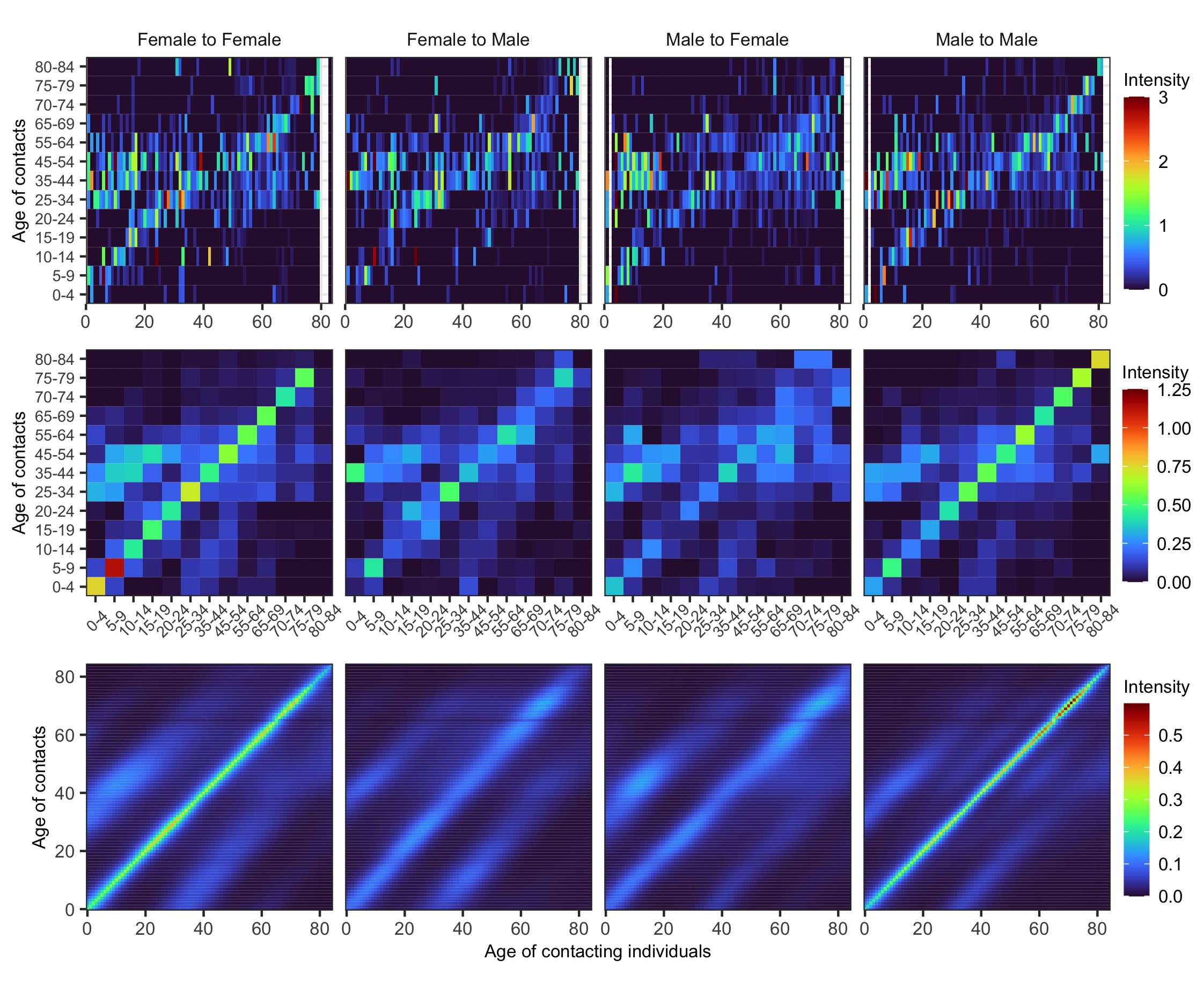}
    \begin{adjustwidth}{0.75in}{0in}
        \captionsetup{width=\linewidth}
        \caption*{{\bf S9 Fig. Empirical and estimated contact intensity patterns for COVIMOD wave 5}. (Top row) Crude empirical social contact intensity patterns, with crude contact intensities above a value of 3 truncated for visualisation purposes. There are some age groups with no participants, and they are represented by white vertical columns. (Middle row) Contact intensity patterns as estimated by the \texttt{socialmixr} R package~\cite{funk_socialmixr_2020}. (Bottom row) Contact intensity patterns are given by our Bayesian model. The exact runtime arguments for this comparison are given in script \texttt{figure-6-9.R} on our accompanying GitHub repository.}
    \end{adjustwidth}
\end{adjustwidth}
\end{figure}

\begin{figure}[!t]
\begin{adjustwidth}{-2.25in}{0in}
    \includegraphics[width=\linewidth]{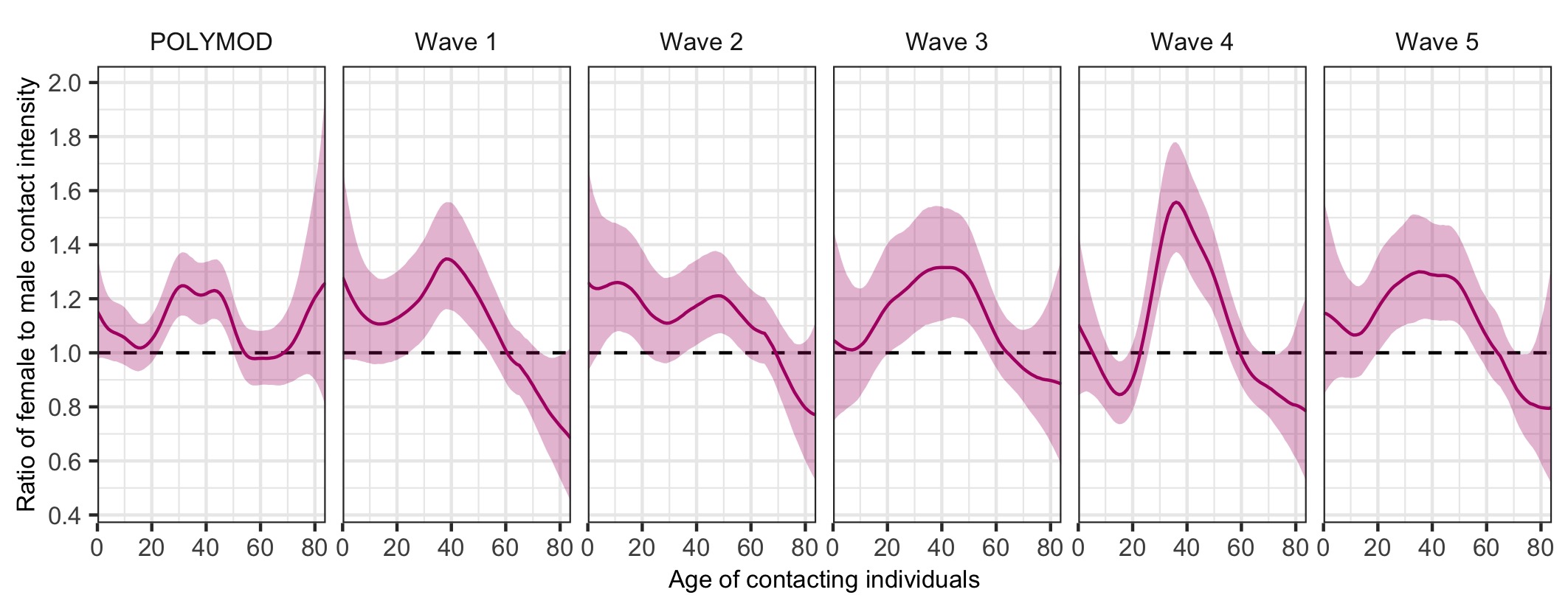}
    \begin{adjustwidth}{0.75in}{0in}
        \captionsetup{width=\linewidth}
        \caption*{{\bf S10 Fig. Ratio of female to male marginal contact intensities for POLYMOD and COVIMOD}. Lines represent posterior median estimates of the ratio of female to male marginal contact intensities, i.e., $m_a^F / m_a^M$. A ratio of 1 (dashed lines) indicates no difference in contact intensities between genders. Shaded ribbons represent 95\% credible intervals.}
    \end{adjustwidth}
\end{adjustwidth}
\end{figure}

\end{document}